\documentclass{article}    
\usepackage{lscape,epsfig,graphicx,subcaption}

\usepackage[section]{placeins}
\usepackage{makecell}
\usepackage{multirow}
\usepackage{siunitx,amssymb}
\usepackage{authblk}
\usepackage{color}
\usepackage{url}
\usepackage{overpic}
\usepackage {lineno}
\title{\textbf{Design of the PMT underwater cascade implosion protection system for JUNO}}
\author[a]{Miao He\thanks{hem@ihep.ac.cn}}
\author[a]{Zhonghua Qin}
\author[a]{Shaojing Hou}
\author[a]{Xiaoping Jing}
\author[c]{Hongbang Liu}
\author[a]{Zunjian Ke}
\author[a,b]{Diru Wu}
\author[a]{Wan Xie}
\author[a]{Mehang Xu}
\author[a]{Fang Chen}
\author[a]{Junguang Lu}
\author[a]{Yuekun Heng}
\author[a]{Jiawen Zhang}
\author[a]{Xiaoyan Ma}
\author[d]{Zhipeng Du}
\affil[a]{Institute of High Energy Physics, Beijing 100049, China}
\affil[b]{University of Chinese Academy of Sciences, Beijing 100049, China.}
\affil[c]{Guangxi University, Nanning 530004, China}
\affil[d]{Naval Research Institute, Beijing 100161, China}

\begin{document}
\maketitle
\begin{abstract}
  Photomultiplier tubes (PMTs) are widely used underwater in large-scale neutrino experiments. As a hollow glass spherelike structure, implosion is unavoidable during long-term operation under large water pressure. There is a possibility of cascade implosion to neighbor PMTs due to shockwave. Jiangmen Underground Neutrino Observatory designed a protection structure for each 20-inch PMT, consisting of a top cover, a bottom cover, and their connection. This paper introduces the requirement and design of the PMT protection system, including the material selection, investigation of manufacture technology, and prototyping. Optimization and validation by simulation and underwater experiments are also presented.
\end{abstract}
\section{Introduction}

Jiangmen Underground Neutrino Observatory (JUNO) is designed for the determination of the neutrino mass ordering, precision measurement of neutrino oscillation parameters, and other physics motivations~\cite{An:2015jdp}. The JUNO detector consists of 20 kiloton liquid scintillator and more than 40 thousand photomultiplier tubes (PMTs), immersed in a 44~m deep water pool~\cite{Djurcic:2015vqa, JUNO:2022hxd}. The expected energy resolution is 3\% @ 1~MeV to precisely measure the fast oscillation structure in the reactor antineutrino spectrum. The lifetime of JUNO is expected to be larger than 20~years.

All PMTs used in JUNO were made of quartz glass evacuated to $\sim10^{-5}$~Pa. Therefore, pressure coming from water and the atmosphere on the PMT is up to 0.54~MPa. The large hydrostatic pressure may break a PMT glass bulb during long-term operation. In this case, shockwave will be generated because of water rushing into the vacuum and then reversely ejecting in a short time. It propagates in water and creates cascade implosion to a large number of PMTs as a chain reaction, which happened in Super-Kamiokande in 2001~\cite{Super-Kamiokande:2002weg}. Super-Kamiokande rebuilt the full detector including all $\sim$11,000 20-inch PMTs in 2006 with a protection structure assembled to each PMT, made of acrylic on the top and fiberglass-reinforced plastic on the bottom~\cite{Super-Kamiokande:2010tar}. This structure will largely reduce the water flow in case of implosion of a single PMT, and thus prevent the generation or significantly reduce the strength of the shockwave to avoid the chain reaction. This design has been inherited by Hyper-Kamiokande~\cite{Hyper-Kamiokande:2018ofw}, while fiberglass is going to be replaced with stainless steel for less light emission and radioisotopes.

Simulation and experimental studies for the 12-inch PMT underwater implosion and the impact on the neighbor PMTs were carried out by LBNE~\cite{LBNE:2012ute, Diwan:2012zz, Ling:2013xea}. The energy of the shockwave is in proportion to the product of water pressure and the PMT volume, while the peak pressure at a certain distance $L$ from the implosion center drops as $\sim L^{-1}$. Similar studies done by JUNO in 2015~\cite{juno_implosion} using 20-inch PMTs in 0.5~MPa water showed the peak pressure to the neighbor PMT surface was about 20~MPa, and the chain reaction always happened if there is no protection to the imploded PMT. It took about 10~ms for the shockwave to generate since the PMT started crushing. The propagation speed of shockwave in water was measured as 1481~m/s like the sound and the magnitude dropped around $L^{-1.2}$, the latter depending on the shape and crushing time of the glass bulb, as well as neighbor PMTs and other mechanical structure.

The water depth in JUNO is similar to Super-Kamiokande, while the number of 20-inch PMTs in JUNO is almost twice of Super-Kamiokande and they are more closely packed. Therefore it is more challenging to deploy the protection structure. Both design and manufacture need to be reviewed. In addition, transparency of the protection structure in front of the PMT photocathode is more critical for JUNO since it affects the detected number of photons and then the energy resolution. In this paper, we will present the R\&D of the protection system in JUNO. The requirement will be introduced in Sec.~\ref{sec.req}. Design and optimization are presented in Sec.~\ref{sec.design}, including the material selection, manufacturing technology investigation, prototyping, underwater tests, and simulation. The final design and validation are discussed in Sec.~\ref{sec:validation}. Sec.~\ref{sec.summary} is the summary and perspective.

\section{JUNO requirements on implosion protection}
\label{sec.req}

\begin{figure}[ht]
\centering
\begin{subfigure}{.45\textwidth}
  \centering
  \includegraphics[height=0.8\textwidth]{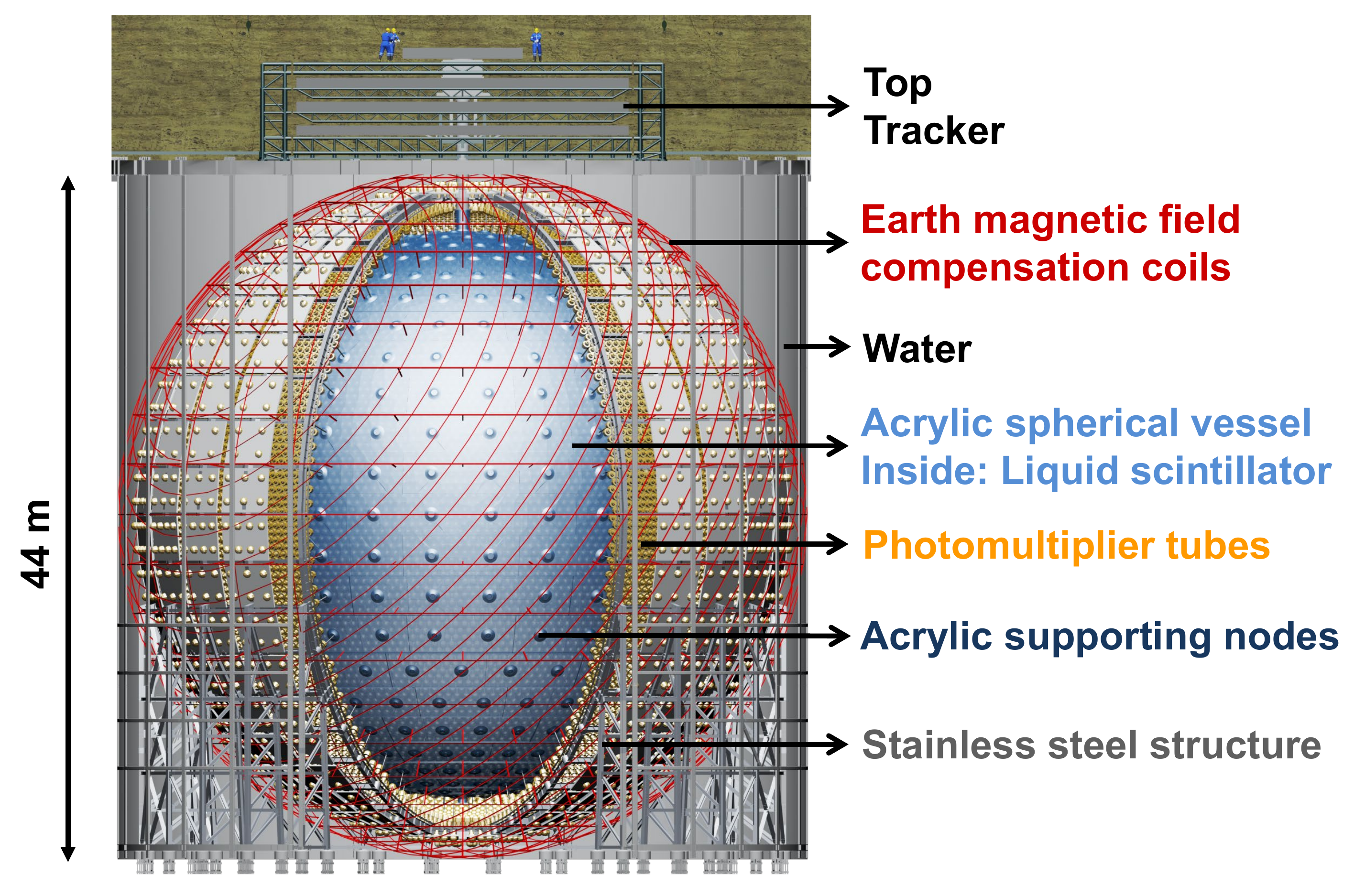}
  \caption{}
  \label{fig:juno.detector1}
\end{subfigure}
\begin{subfigure}{.45\textwidth}
  \flushright
  \includegraphics[height=0.8\textwidth]{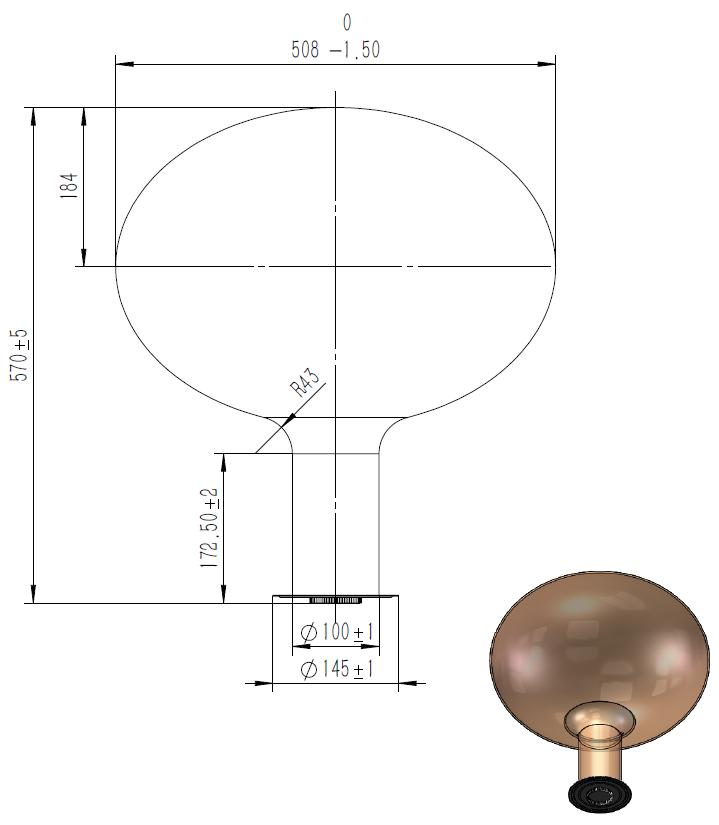}
  \caption{}
  \label{fig:juno.detector2}
\end{subfigure}
\caption{(a) Schematic drawing of the JUNO detector. 20 kiloton liquid scintillator is contained in an acrylic spherical vessel which is supported by a stainless steel truss through 590 supporting nodes. 20,012 20-inch PMTs and 25,600 3-inch PMTs are installed on the stainless steel structure and are surrounded by earth magnetic field compensation coils. All of the detector components are immersed in a 44-m high water pool. A plastic-scintillator based detector is on top of the water pool to track cosmic muons. (b) Schematic drawing of the 20-inch PMT from North Night Vision Technology Co. Ltd. }
\label{fig:juno.detector}
\end{figure}

The design of the JUNO detector is shown in Fig.~\ref{fig:juno.detector}~(a). Liquid scintillator is filled in a 12-cm thick acrylic vessel with an inner diameter of 35.4~m, which is supported by a stainless steel structure through 590 connecting nodes. The majority of scintillation light is between 380~nm and 550~nm, with the most probable value around 420~nm~\cite{JUNO:2020bcl}. 17,612 20-inch PMTs and 25,600 3-inch PMTs are installed surrounding the liquid scintillator, providing 77.9\% optical coverage. All the detector is in a water pool, which also serves as a water Cherenkov detector after equipping with 2,400 20-inch PMTs to tag cosmic muons. There is another plastic scintillator detector on top of the water pool providing precision tracking of cosmic muons.

The total number of $\sim$20,000 20-inch PMTs consists of two types: $\sim$15,000 Micro-Channel Plate (MCP) PMTs from North Night Vision Technology Co., Ltd (NNVT) and $\sim$5,000 dynode PMTs from Hamamatsu Photonics K.K. (HPK). Taking the NNVT PMT as an example (Fig.~\ref{fig:juno.detector}~(b)), the glass bulb is made of an $\sim$3~mm thick ellipsoid with a neck. The total volume in the glass bulb is $\sim$50~L. The HPK PMT has a similar shape and dimensions but the neck is larger, resulting in $\sim$10~L more volume.


The minimum clearance between two 20-inch PMTs is only 25~mm in JUNO to achieve the required optical coverage, which is almost an order of magnitude smaller compared to Super-Kamiokande (40\% coverage~\cite{Super-Kamiokande:2002weg} and $\sim$20~cm clearance). This is the major constraint for JUNO that the thickness of the protection structure needs to be optimized and precisely controlled. This small clearance also results in a factor of two larger peak pressure on the neighbor PMT surface between JUNO and Super-Kamiokande. The JUNO requirements on cascade implosion protection are as follows:
\begin{itemize}
  \item 50~m pedestal water pressure resistant. Unlike the hydrostatic pressure on the PMT, there will be pedestal pressure on the protection structure when the inside PMT implodes. 
  \item No chain reaction in case of a single PMT implosion. Hydrostatic water pressure tolerance of the JUNO 20-inch PMT is about 1~MPa~\cite{JUNO:2022hlz}, and we assume the implosion threshold from shockwave is larger than 1~MPa because its impact time is only tens of microseconds.
  \item Accommodate 25~mm clearance.
  \item Light absorption $<$~1\% at 420~nm and $<$7\% at 380~nm. Taking acrylic as an example, these requirements correspond to $>$91\% and $>$85\% transparency in air at 420~nm and 380~nm, respectively, where there is an approximate 8\% loss of light due to reflection between acrylic and air. In water, reflected light will be smaller than 1\% because of the closer refractive index to air.
  \item Compatible with pure water.
  \item Low background. At least an order of magnitude lower radioactivity compared with the PMT itself.
\end{itemize}

\section{Preliminary design and manufacture investigation}
\label{sec.design}

\begin{figure}[!hbt]
  \centering  
  \includegraphics[width=0.8\textwidth]{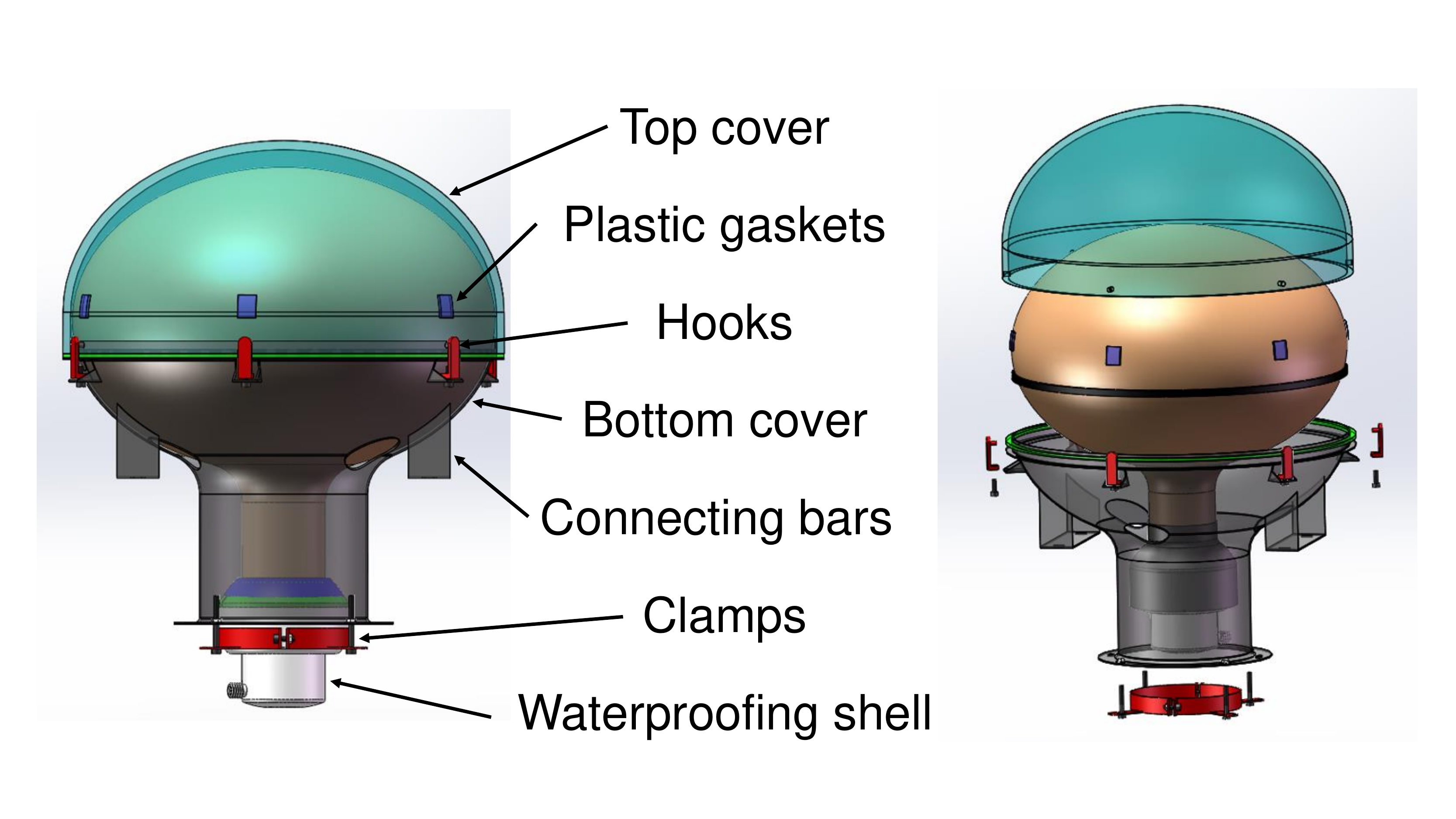}
  \caption{Conceptual design of the PMT protection system. The top cover is transparent and the bottom cover is made of stainless steel. The two covers are connected by six hooks. Several plastic gaskets are used as buffer between the PMT glass and the protection cover. Another stainless steel shell filled with sealants is on the bottom of the PMT for the waterproofing and is fixed by a pair of clamps. The full PMT unit is installed on the detector main structure through two connecting bars.}
  \label{fig:conceptual.design}
\end{figure}

The conceptual design of the JUNO PMT protection is similar to Super-Kamiokande. Each PMT will be equipped with a pair of covers, consisting of a top cover and a bottom cover, as shown in Fig.~\ref{fig:conceptual.design}. The top cover is a shell combined with a semi-ellipsoid and a short cylinder. It is transparent, with the material investigated and selected from several candidates. The bottom cover is a truncated ellipsoid connected to a cylinder, made of stainless steel. There are a few holes on each cover (not shown in the figure) to let water in and air out. Six hooks or screws connect the two covers. A stainless steel shell filled with sealants provides waterproofing for the electrical components on the bottom of the PMT, and this shell is fixed by a pair of clamps. There are several plastic gaskets between the PMT and two covers for support. There are two connecting bars on the stainless steel cover serving as the interface to the detector's main structure. In this paper, we only introduce the design of the top cover, the bottom cover, and the connection structure in between, while other waterproofing or mechanical structure will be reported in the future. In addition, the design was done based on the NNVT PMT and was assumed to be applicable to the HPK PMT.

\subsection{Material selection for the top cover}

Polymethylmethacrylate (PMMA), also known as acrylic, is a transparent thermoplastic widely used in neutrino experiments as a container of the detector target, either water such as SNO~\cite{SNO:1999crp}, or organic liquid such Daya Bay~\cite{DayaBay:2015kir}, Double Chooz~\cite{DoubleChooz:2011ymz}, and RENO~\cite{RENO:2012mkc}. In particular, Super-Kamiokande used acrylic to produce the PMT protection cover after the accident. The water compatibility was found to be excellent and the radioactivity can be easily controlled to be better than 10~ppt~\cite{SNO:1999crp, JUNO:2021kxb}. On the other hand, acrylic is fragile and the elongation at break is only a few percent, which means a little deformation could end up with a fracture. Therefore, two other materials with much larger elongation at break, Polycarbonate (PC) and Polyethylene terephthalate (PET) were also suggested. The major mechanical parameters of these three materials were investigated and compared in Table~\ref{tab:material}.

\begin{table}
  \centering
  \begin{tabular}{ccccc}
    \hline
    Material & PMMA & PC & PET \\
    \hline
    Refractive index & 1.49 & 1.58 & 1.58-1.64 \\
    Elongation at break (\%) & 2.5-4 & 100-150 & 60-165 \\
    Tensile strength (MPa) & 80 & 55-75 & 80 \\
    Tensile modulus (MPa) & 2400-3300 & 2300-2400 & 2000-4000 \\
    Izod impact strength (J$\cdot$m$^{-1}$) & 16-32 & 600-850 & 13-35 \\
    \hline
	\end{tabular}
	\caption{Comparison of transparent material for the top cover. All data are taken from Ref.~\cite{material-website} for consistency. It is worth to note that mechanical performances also depend on the manufacture technology. In particular, tensile strength of PMMA shown here corresponds to casting, while injection molding typically gives lower result similar to PC, which will be shown in Table~\ref{tab:PMMA}.}
	\label{tab:material}
\end{table}

\begin{figure}[ht]
\begin{subfigure}{.5\textwidth}
  \centering
  \includegraphics[width=\textwidth]{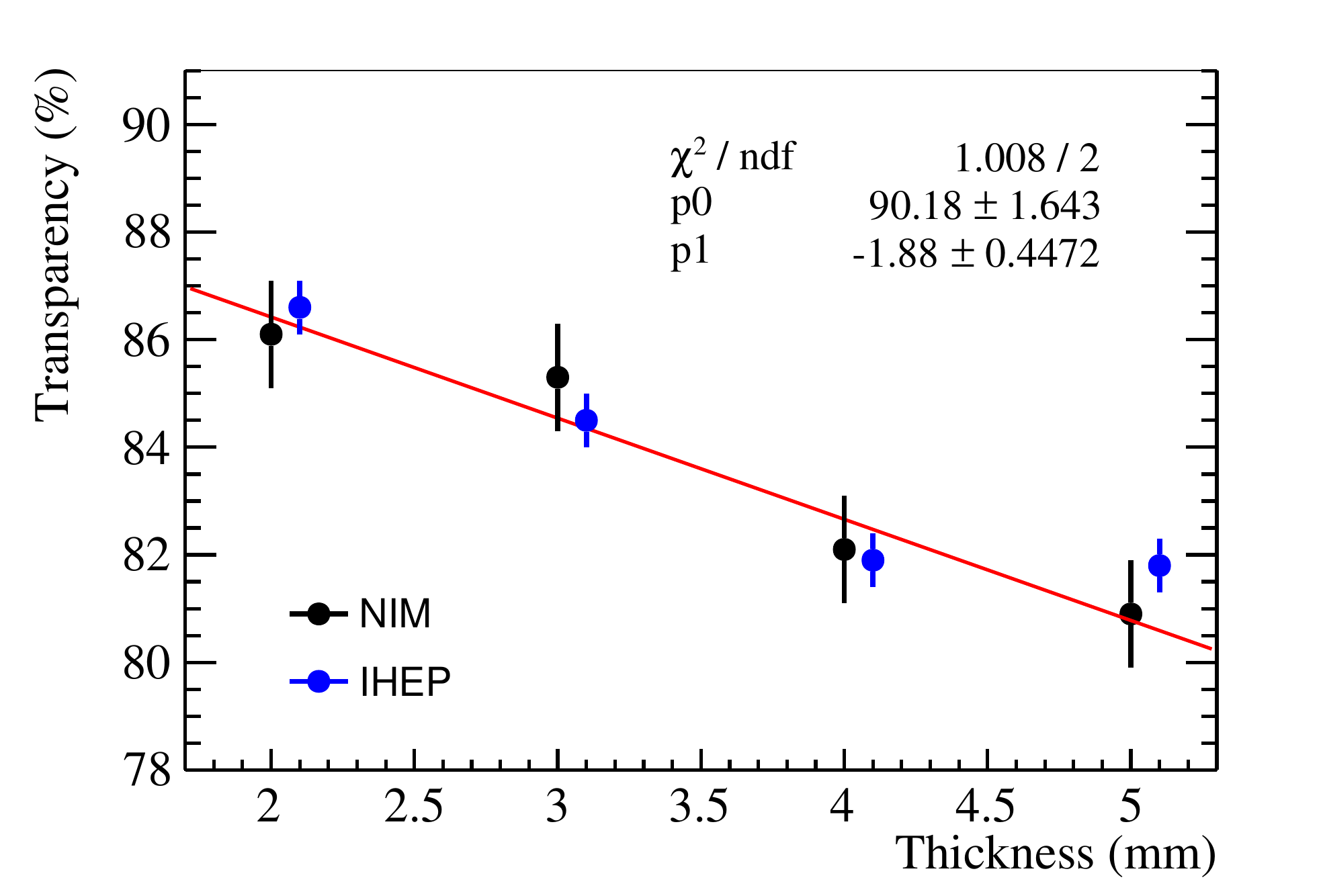}
  \caption{}
\end{subfigure}
\begin{subfigure}{.5\textwidth}
  \centering
  \includegraphics[width=\textwidth]{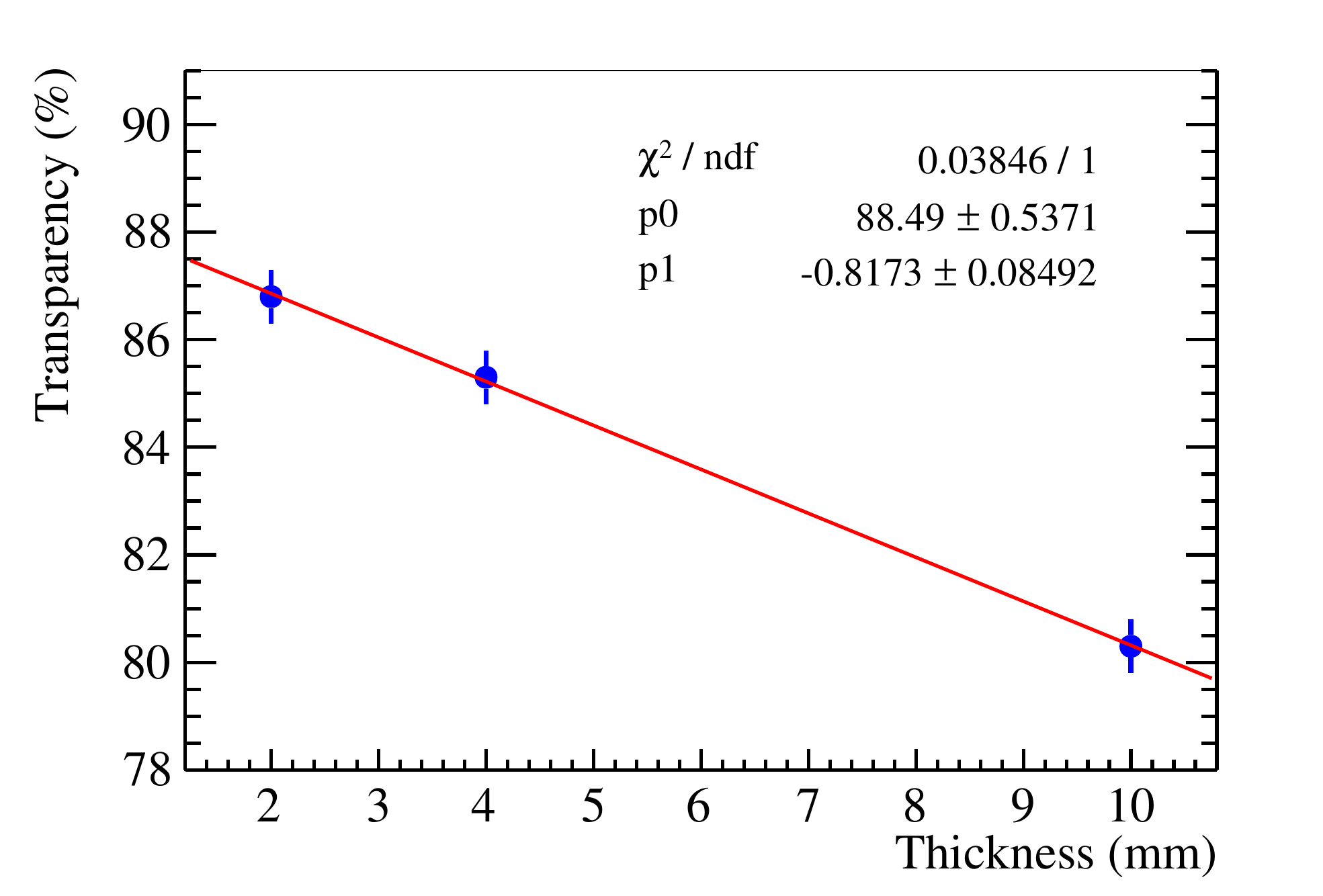}
  \caption{}
\end{subfigure}
\caption{Transparency at 420~nm in air of (a) PET and (b) PC as a function of their thicknesses.}
\label{fig:trp.pc.pet}
\end{figure}

A few PC and PET sheets were made with different thicknesses, and the transparency in air was measured with a commercial spectrophotometer, with the results at 420~nm shown in Fig.~\ref{fig:trp.pc.pet}. The uncertainty was estimated as 0.5\%, by measuring the same sample several times and by comparing it with another transparency measurement system~\cite{Yang:2020juno}. The light absorption was assumed to be linear to such a small thickness and fitted by $y=p0+p1\cdot x$, where $y$ represents the transparency and $x$ represents the sample thickness. The PET samples were also sent to National Institute of Metrology (NIM) and the measurements gave very consistent results. From the fitting, the light absorption was found to be 1.9\% per mm for PET and 0.8\% per mm for PC. Taking into account both the light absorption and the optical coverage, to reach the same level of detected light with 10~mm acrylic cover, the thickness of the PET cover and PC cover can only be 1.5~mm and 3~mm, respectively. Samples of PET covers with 1.5-5~mm thickness and PC covers with 3~mm thickness were produced while all of them failed in the underwater tests, which will be introduced in Sec.~\ref{sec:uwtest}. In the end, acrylic was chosen to be the material for the top cover.

%

\subsection{Manufacture technology investigation and prototyping for the top cover}

Three technologies were tried to produce the acrylic top cover.
\begin{itemize}
  \item \textbf{Casting}: Almost all of the acrylic sheets in the market are made by casting, where methyl methacrylate (MMA) forms a longer molecular chain by the polymerization reaction and turns into PMMA. A dedicated mold was made by a very thick acrylic plate with aluminum film pasted on the surface. It took four days to produce one cover because of the slow polymerization reaction rate. In addition, a large number of bubbles were generated during this process and were hard to be excluded in such kind of mold.
  \item \textbf{Thermoforming}: An acrylic sheet pressurized by a stainless steel mold at a temperature between 100 and 150~$^{\circ}$C. The process took a few hours. More than 10 samples were produced. The thickness was measured by an ultrasonic device at different positions and the deviation was found to be larger than 50\%.
  \item \textbf{Injection molding}: Dry acrylic pellets were fed into an injection machine. At a temperature of about 240~$^{\circ}$C, they became semifluid and were injected into a steel mold at a temperature of 70~$^{\circ}$C and a successively decreased pressure from 140~MPa to 60~MPa. After several minutes, it became solid and can be removed from the mold.
\end{itemize}

\begin{figure}[!hbt]
  \centering
  \includegraphics[width=0.7\textwidth]{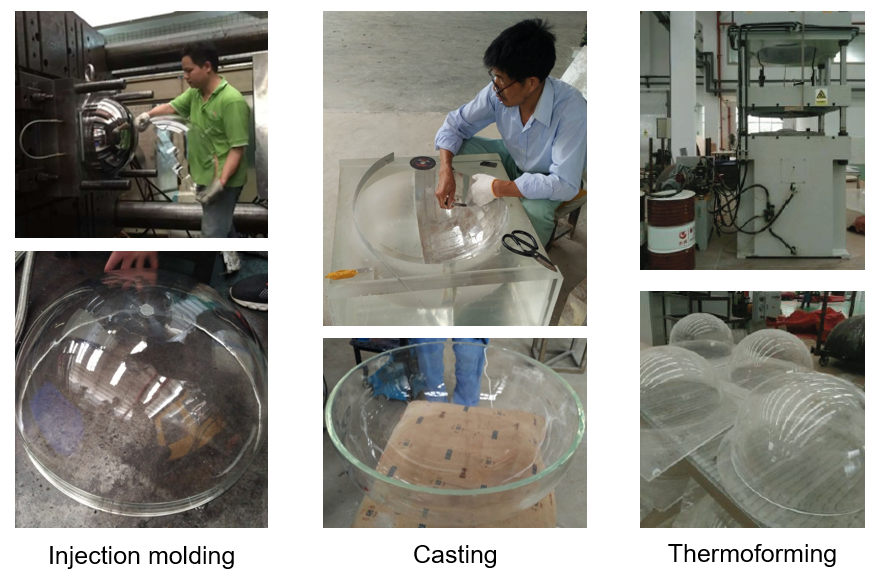}
  \caption{Manufacture of acrylic covers by three different technologies.}
  \label{fig:technology}
\end{figure}

The molds and products for all of these three technologies are shown in Fig.~\ref{fig:technology}. Injection molding was found to be the best for the acrylic cover because of the high production rate and precise control of the dimensions with a deviation smaller than 1~mm.

\subsection{Study of injection molding}
\label{sec:injection}
Many companies are providing PMMA raw materials for injection molding. Most of them carry anti-UV components and thus can not be used for the protection cover. Therefore, material screening was done firstly by only selecting types that are explicitly labeled as UV-transmission in their specifications. 13 types of materials were selected with their major mechanical parameters summarized in Table~\ref{tab:PMMA}. The transparencies for all of them are labeled as larger than 92\% in air for 3.2~mm thick standard specimens.

\begin{table}
  \centering
  \begin{tabular}{cccccc}
    \hline
    Company & Brand & Type & Tensile & Tensile & Elongation \\
     &  &  &  strength (MPa) &  modulus (MPa) & at break (\%) \\
    \hline
    \multirow{2}*{Evonik~\cite{Evonik}} & \multirow{2}*{PLEXIGLAS$^\circledR$} & IM20 & 77 & 3300 & 5.5\% \\
     & & 7N & 73 & 3200 & 3.5\% \\
    \hline
     & \multirow{3}*{PLEXIGLAS$^\circledR$} & V920-UVT & 69 & 3100 & 5\% \\
     Altuglas & & VS-UVT & 65 & 2900 & 5\% \\
     International~\cite{Altuglas1,Altuglas2} & & V825-UVA5A & 70 & 3100 & 5\% \\
     & ALTUGLAS$^\circledR$ & VSUVT & 66 & 2900 & 5\% \\
    \hline
    \multirow{3}*{MITSUBISHI~\cite{MITSUBISHI}} & \multirow{3}*{ACRYPET$^\circledR$} & VH5 & 61 & 3300 & 3\% \\
     & & TF8 & 59 & 3300 & 3\% \\
     & & TF9 & 57 & 3200 & 2\% \\
    \hline
    Chimei~\cite{Chimei} & ACRYREX$^\circledR$ & CM211 & 65 & 2700 & 5.5\% \\
    \hline
    \multirow{3}*{Sumitomo~\cite{Sumitomo}} & \multirow{3}*{SUMIPEX$^\circledR$} & MG5 & 75 & 3100 & 3\% \\
     & & MGSS & 73 & 3100 & 2\% \\
     & & MGSV & 70 & 3100 & 2\% \\
    \hline
	\end{tabular}
    \caption{Mechanical specifications of different UV-transmission PMMA material. All data come from companies official websites.}
	\label{tab:PMMA}
\end{table}

\begin{figure}[!hbt]
  \centering
  \includegraphics[width=0.45\textwidth]{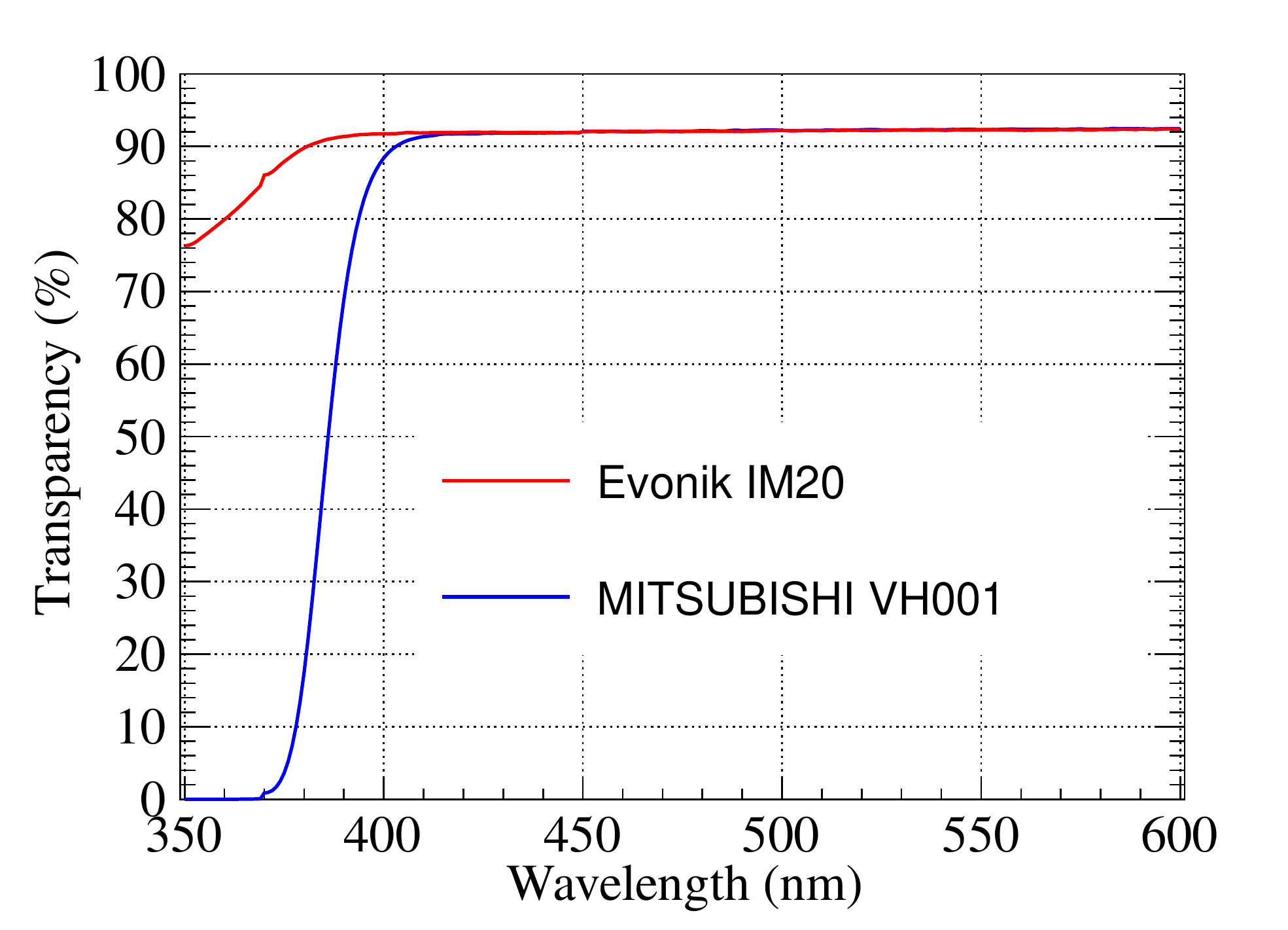}
  \caption{Examples of transparency scanning results in air. IM20 from Evonik is UV-transparent while VH001 from MITSUBISHI contains UV absorber.}
  \label{fig:tp_curve}
\end{figure}

Since UV transmission is critical for JUNO, materials from Table~\ref{tab:PMMA} were purchased from the market as many as possible. 3-mm thick acrylic plates were made with them and the transparency as a function of wavelength was measured by the spectrophotometer. One of the results is shown in Fig.~\ref{fig:tp_curve} where the sample was made by Evonik IM20 and the transparency is 91.9\% at 420~nm and 89.8\% at 380~nm. To demonstrate the UV-absorption, another sample's result is also shown which was made by MITSUBISHI VH001 and contains a UV absorber therefore the transparency decreases quickly from 91.7\% at 420~nm to 17.5\% at 380~nm. Taking into account the mechanical performances and price, two materials, IM20 from Evonik and MGSV from Sumitomo were selected as candidates. Both of them have transparency around 92.0\% at 420~nm, and have been used to produce acrylic cover prototypes and tested successfully, which will be introduced in Sec.~\ref{sec:exp_val}.

Hygroscopicity is one of the properties of acrylic. Water absorption can be considered as diffusion with the coefficient changes with the temperature~\cite{water_abs1}. However, the mass fraction of absorbed water at the equilibrium state does not depend on the volume or shape of acrylic, nor the temperature either. It was measured around 2\%~\cite{water_abs2, water_abs3}. Some studies suggested that water as a polar molecule has a plasticization effect and may introduce some stress on the acrylic thus reducing the mechanical performances~\cite{water_abs4}.

\begin{figure}[!hbt]
  \centering
  \includegraphics[width=0.5\textwidth]{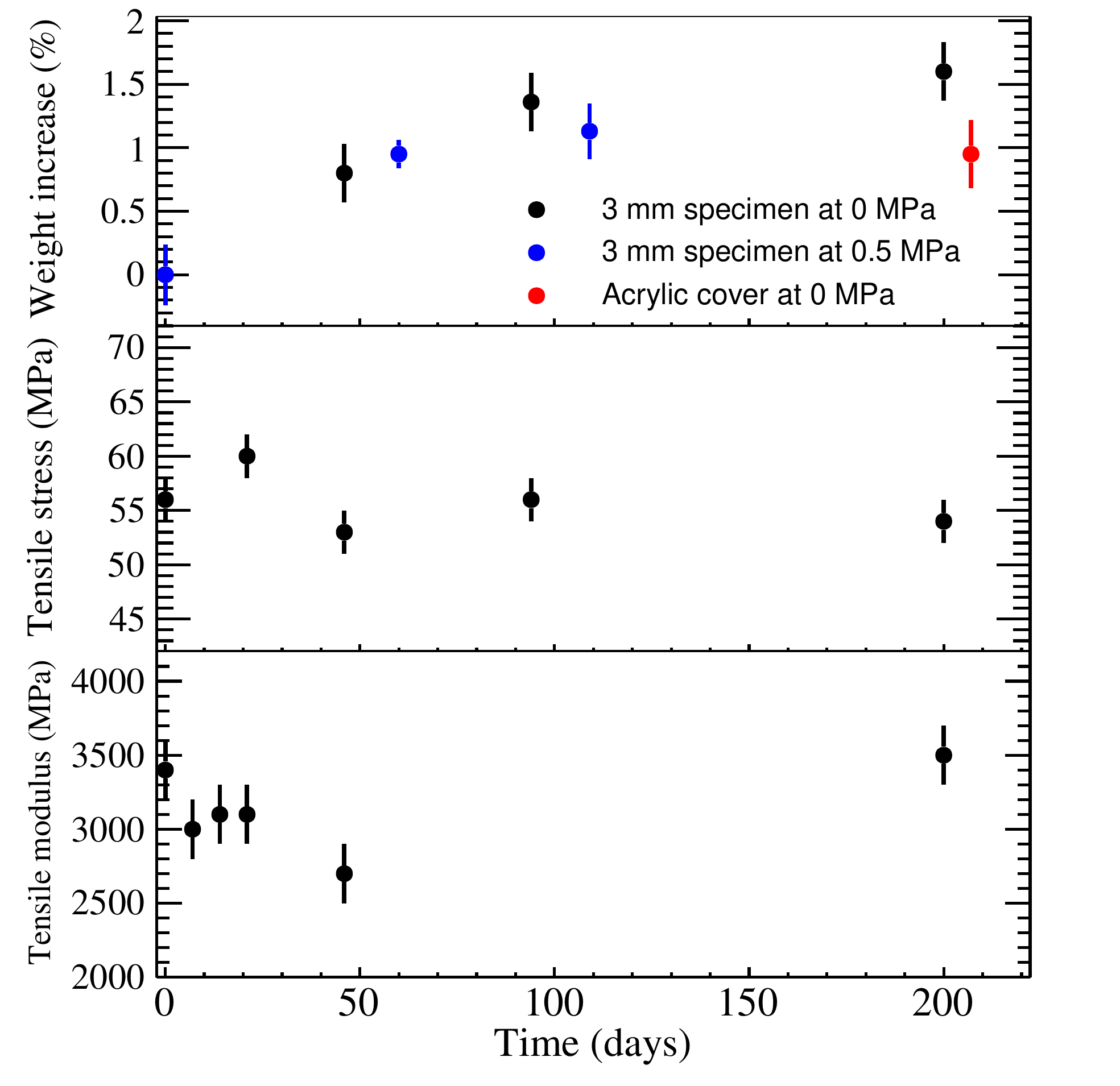}
  \caption{The top panel is the weight increase due to water absorption as a function of time for different acrylic samples and water pressures. The middle panel and the bottom panel show the change of tensile strength and tensile modulus of acrylic specimens as a function of time in water under normal pressure.}
  \label{fig:water_abs}
\end{figure}

A batch of standard specimens with the shape following ISO~527 were injection molded with an arbitrary PMMA material. Their weight was measured as 11.54~g with a standard deviation of 0.02~g, then some of them were put into ultra-pure water at normal pressure and the rest into tap water at 0.5~MPa pressure, both at room temperature. In the following 200 days, several of them were taken out and measured again. The increase in the weight represented the water absorption as shown in the top panel of Fig.~\ref{fig:water_abs}. The absorption rates were very similar for different water and different pressures. The water mass fraction was 1.60\% after 200 days and was close to saturation. The weight of two acrylic covers was also measured before and after 207 days in tap water at normal pressure and room temperature. The absorption rate was smaller than the standard specimen because of its larger thickness.

The tensile stress and tensile modulus were also measured for these specimens with the standard GB/T 1040.2-2006, as shown in the middle and the bottom panels of Fig.~\ref{fig:water_abs}. No decrease for either of the two parameters was found during these 200 days. Since the water absorption in our experiment was already 80\% compared to the saturation, a significant reduction in the mechanical performance is not expected. On the other hand, during the long-term operation of JUNO, the protection cover will work in the environment almost without any UV light at a stable temperature 21$\pm$1~$^{\circ}$C, and the protection structure will not suffer any external stress except for its weight in water, the aging should not be a problem either.

\subsection{Investigation and prototyping of the bottom cover}

The bottom cover is used to support the top cover and to provide an interface to the detector's main structure. Stainless steel type 304 was chosen because of many advantages:
1) commonly used material in water with low background easily to be $\sim$ppb level;
2) good mechanical performance with a large ultimate strength of more than 500~MPa;
3) good processability with many techniques;
4) compatible with pure water;
5) low cost compared to stainless steel type 316 with similar mechanical performance.

\begin{figure}[ht]
\begin{subfigure}{.5\textwidth}
  \centering
  \includegraphics[height=0.6\textwidth]{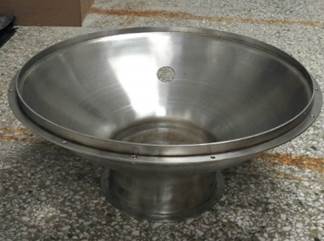}
  \caption{}
\end{subfigure}
\begin{subfigure}{.5\textwidth}
  \centering
  \includegraphics[height=0.6\textwidth]{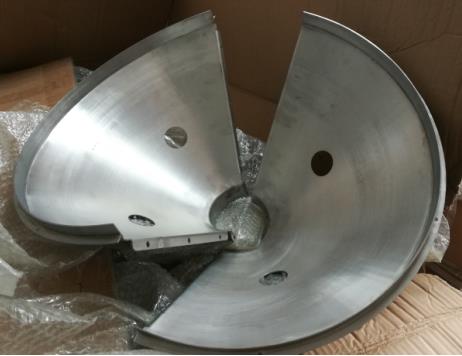}
  \caption{}
\end{subfigure}
\caption{Bottom cover prototypes made by two different technologies including (a) stamping and (b) rolling.}
\label{fig:sscover}
\end{figure}

Different technologies are investigated for producing the bottom cover, including rolling, stamping, rotary extrusion, etc. A prototype with a semi-ellipsoid shape was made by stamping as shown in Fig.~\ref{fig:sscover}~(a) and another one with a cone shape by rolling in Fig.~\ref{fig:sscover}~(b). Rolling is cheaper for prototyping since it does not require a mold. However, it is very difficult to produce an ellipsoid by rolling even splitting the bottom cover into two halves. On the other hand, the cone-shaped bottom cover shows significantly lower mechanical performance according to the simulation study in Sec.~\ref{sec:sim}. Therefore, stamping was used for the main body and the rest connecting structure was welded. Demagnetization after stamping is needed since there will be a magnetic field because of stainless steel stretching, which has a significant impact on the photoelectrons collection in the PMT.

\subsection{Underwater tests}
\label{sec:uwtest}

\begin{figure}[!hbt]
  \centering
  \includegraphics[width=0.7\textwidth]{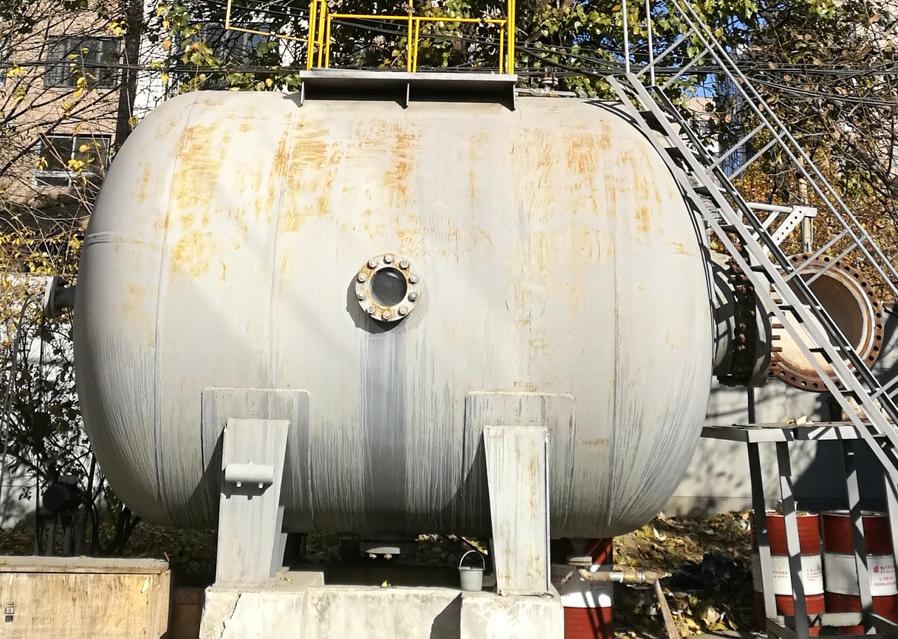}
  \caption{Pressure tank made of steel used for the underwater test. The inner volume is 35~m$^3$. Four acrylic windows on the wall of the tank are used for the high-speed camera or the light.}
  \label{fig:pressure_tank}
\end{figure}

The underwater experiments were designed to test the full PMT protection system. A steel-made water tank was produced to provide up to 1~MPa water pressure. The tank looks like a horizontal located cylinder, as shown in Fig.~\ref{fig:pressure_tank}, with an inner diameter of 3~m and a length of 5~m. During the experiments, water was injected to a level of 4/5 and the rest 1/5 was kept in air. After that, the pressure was increased with the help of an air pump, typically stopped at 0.5~MPa with respect to the atmospheric pressure. Since the air volume is two orders of magnitude larger than a PMT (typically 0.05~m$^3$), the decrease in the water pressure is negligible even if several PMTs break at the same time. There are four acrylic windows on the tank allowing recording of the experiment by the high-speed camera with the help of high-intensity light out of the tank.

\begin{figure}[!hbt]
  \centering
  \includegraphics[width=0.45\textwidth]{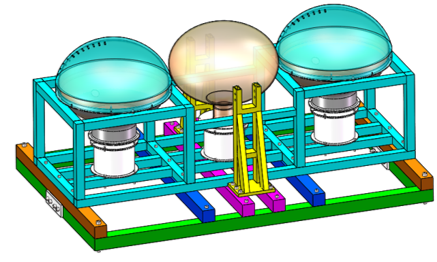}
  \includegraphics[width=0.45\textwidth]{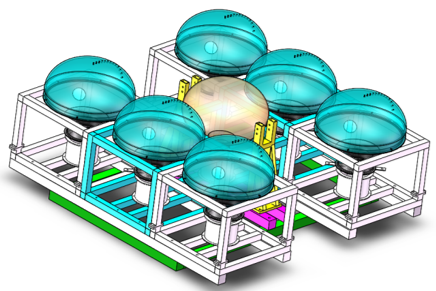}
  \caption{Setup of the PMT underwater protection experiment at the optimization stage. The central PMT without any protection cover plays as a trigger to implode neighbor PMTs all equipped with the protection structure.}
  \label{fig:pmt_module}
\end{figure}

In the traditional way like Super-Kamiokande, a PMT with protection covers was imploded underwater and the neighbor PMTs were used to justify where there was a chain reaction. In this scenario, only one pair of top and bottom covers can be tested at a time. To increase the test efficiency, a new setup was implemented as shown in Fig.~\ref{fig:pmt_module}. A 3$\times$3 PMT module was produced of stainless steel. The central PMT was installed without any protection cover or only with the bottom cover. Two 14$\times$14~cm$^2$  stainless steel plates with four screws at the corner were installed around the equator of the PMT face to face, driven by a hydraulic device to implode the PMT underwater. The shockwave was therefore generated and used as a trigger to implode neighbor PMTs, and thus their protection covers will stand under a pedestal water pressure. Once they fail, there will be secondary shockwave which can be measured by pressure sensors. With this setup, up to six sets of protection covers can be tested simultaneously.

The shockwave propagates in the water like sound. When it comes to the boundary between water and acrylic or stainless steel, part of the energy is reflected which can be calculated as
\begin{equation}\label{eq:swr}
    r = (\frac{\rho_2c_2-\rho_1c_1}{\rho_2c_2+\rho_1c_1})^2,
\end{equation}
\begin{equation}\label{eq:swt}
    t = 1-r = \frac{4\rho_2c_2\rho_1c_1}{(\rho_2c_2+\rho_1c_1)^2},
\end{equation}
where $\rho_1$, $\rho_2$ represent material densities and $c_1$, $c_2$ represent speed of sound in two materials. $r$ and $t$ are the reflected and transmitted shockwave energy fractions, respectively. Considering water density 1~g/cm$^3$, speed of sound in water 1483~m/s, acrylic density 1.19~g/cm$^3$, speed of sound in acrylic 2730~m/s, and there are two boundaries when the shockwave goes through the acrylic cover, the transmitted energy fraction is calculated as 74\%. It means $\sim$1/4 of the shockwave energy will be reflected by the acrylic cover and possibly bring some damages, thus this experiment is at a very extreme case. The rest of shockwave energy will completely impact on the PMT because there is vacuum inside the PMT thus the shockwave can not propagate further.

\begin{figure}[ht]
\begin{subfigure}{.5\textwidth}
  \centering
  \includegraphics[height=0.5\textwidth]{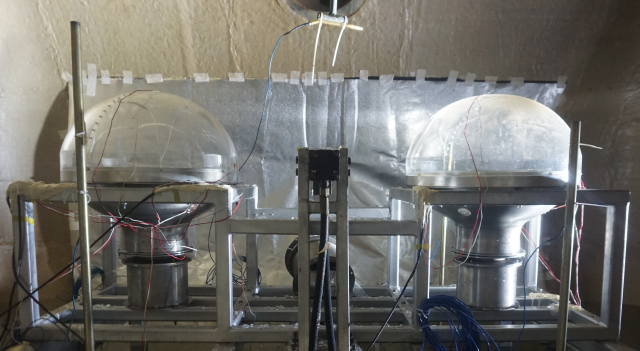}
  \includegraphics[width=0.9\textwidth]{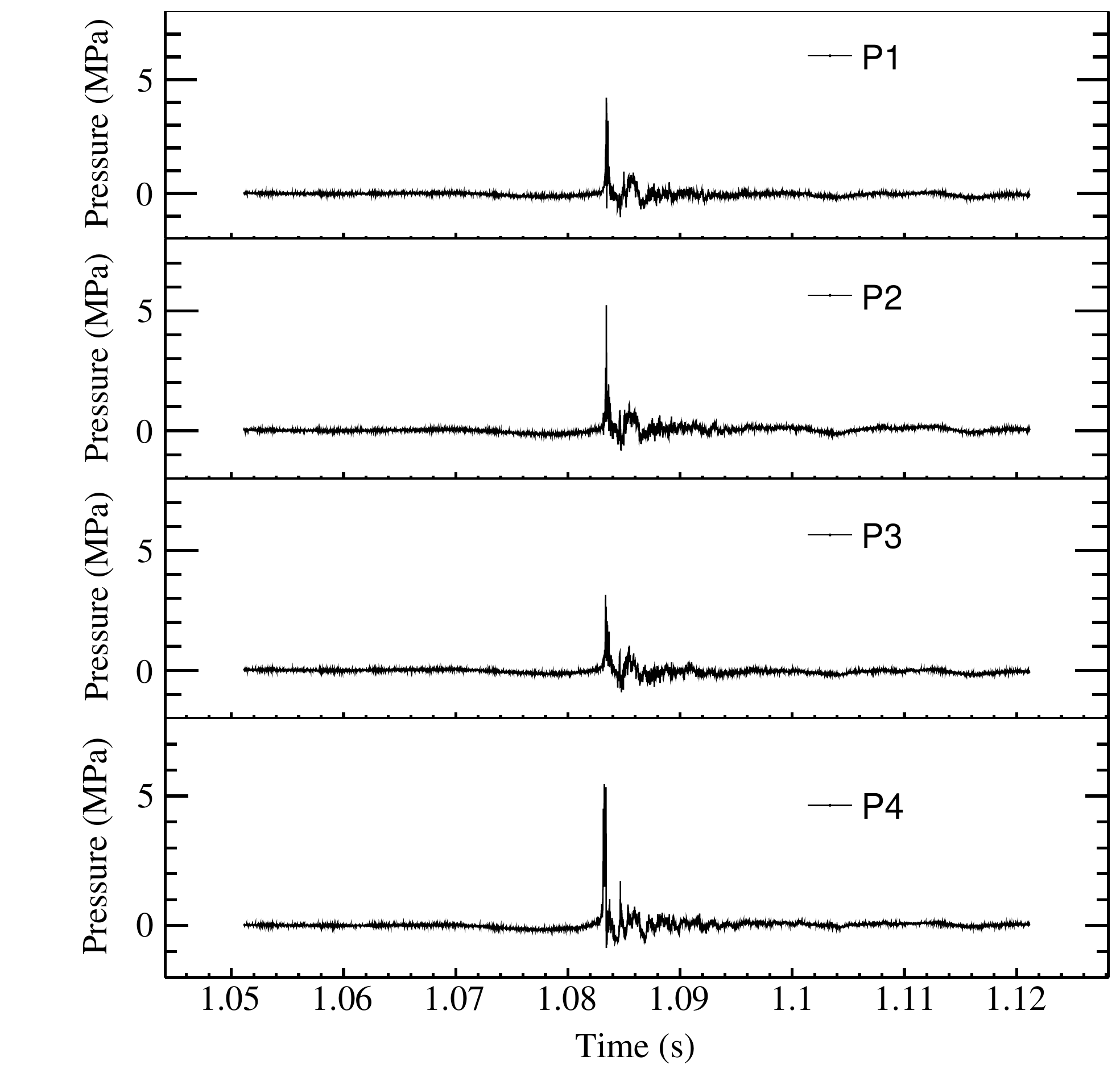}
  \caption{}
  \end{subfigure}
\begin{subfigure}{.5\textwidth}
  \centering
  \includegraphics[height=0.5\textwidth]{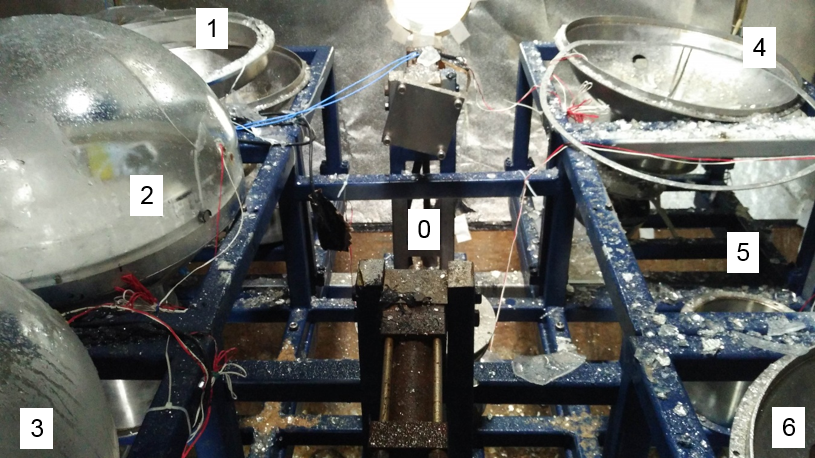}
  \includegraphics[width=0.9\textwidth]{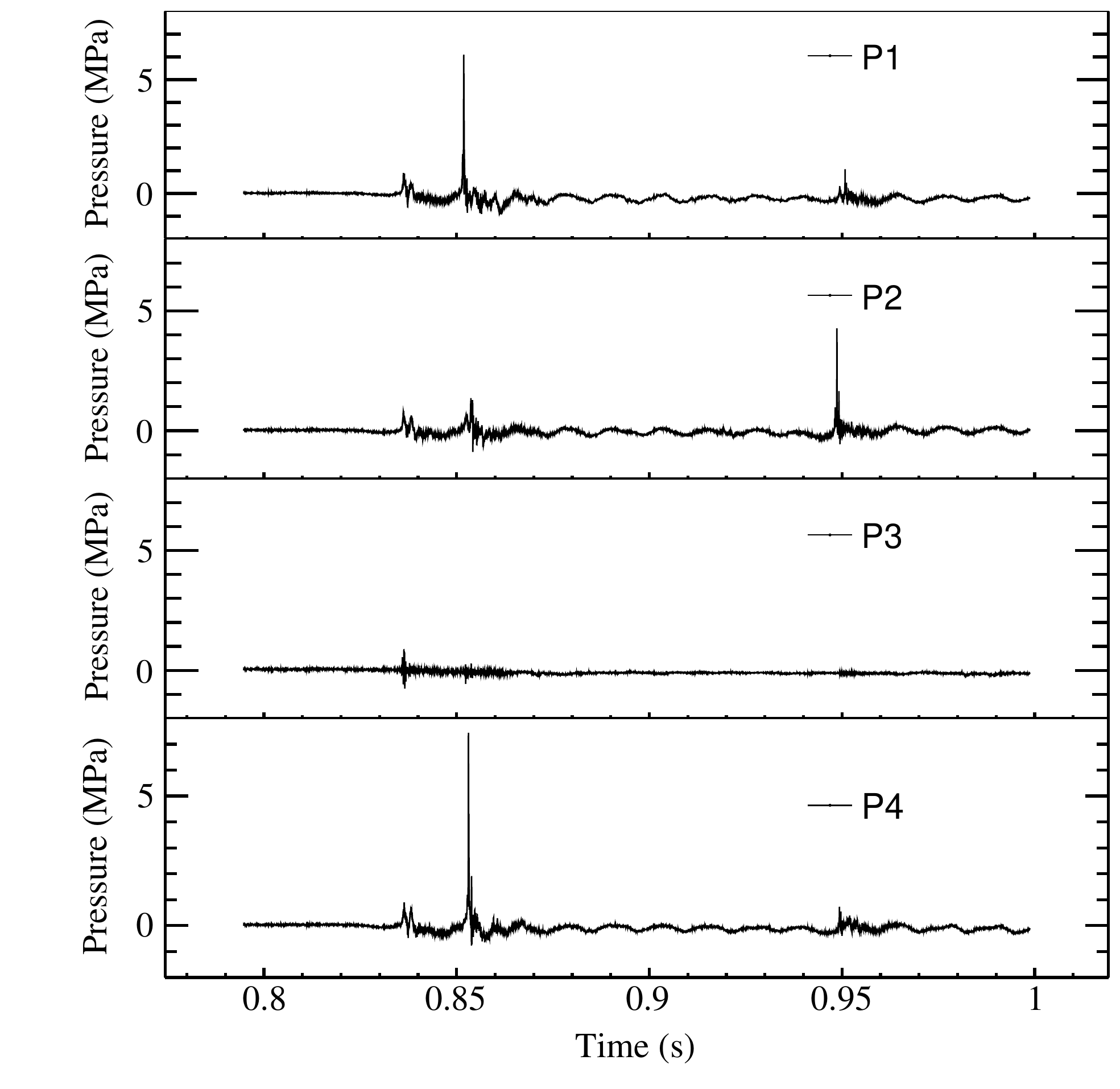}
  \caption{}
  \end{subfigure}
\caption{PMT underwater experiments. The central PMT was a trigger which created shockwave and imploded all neighbor PMTs in protection covers. (a) Two acrylic covers made of injection molding. (b) Six top covers made of different material, manufacture technology and thickness. The bottom panels show the shockwave measurements by four pressure sensors.}
\label{fig:test1}
\end{figure}

Six underwater experiments were done in 2016 and 2017 to find an optimized thickness for the top cover, while the bottom covers were kept the same. Two typical experiments were introduced here. In Fig.~\ref{fig:test1}~(a), two PMTs both with 11.5~mm thick injection molding acrylic covers were installed close to the central trigger PMT. Four pressure sensors were deployed 0.53~m-0.75~m to their nearest PMT centers. After the implosion, the two neighbor PMTs were broken as expected, while their protection covers survived. In particular, there are several cracks on the acrylic covers all starting from the connection screws, which will be discussed in Sec.~\ref{sec:connection}. Pressures of the shockwave as a function of time are shown in the bottom panel of Fig.~\ref{fig:test1}~(a). The peak at 1.083~s corresponds to the implosion of the central PMT. There are small oscillations in the next 10~ms presumably due to multiple reflections of the primary shockwave on the inner surface of the pressure tank. After that, no peak pressure was observed therefore there was no secondary shockwave generated by the neighbor PMTs, and thus the protection requirement was satisfied. More details of this experiment including the high-speed camera data can be found in Ref.~\cite{navy1}.

This result gives two conclusions. First, it approves that once the protection cover does not collapse in case of an implosion of a PMT, there will be almost no shockwave generated thus no chain reaction happens. Second, in the most extreme case that one protection cover collapses together with its inside PMT because of any production defection, there will be a chain reaction but only limited to a small range unless a second protection cover collapses as well. Therefore qualities of the products have to be controlled to make sure the possibility of the second case is negligible.

Another experiment with six neighbor PMTs is shown in the top panel of Fig.~\ref{fig:test1}~(b), where the numbers represent PMTs and their protection covers. Acrylic covers 2 and 5 were made of injection molding, the same as in the previous experiment. 3 and 4 were made of thermoforming, with thicknesses of 8.5~mm and 6.5~mm, respectively. 1 and 6 were made of PC by blow molding~\cite{blow_molding} with a thickness of 3~mm. After the implosion of the central PMT 0, all six neighbor PMTs broke completely. Acrylic covers 2 and 3 stood at their original places with only some cracks. Acrylic cover 5 also survived but down to the ground together with the stainless steel cover thus they can not be seen in this photo, because the connection between the stainless steel cover and the experiment module was not tight enough. Acrylic cover 4 and two PC covers 1 and 6 were all collapsed and broke into pieces.

There are three peaks clearly seen in the pressure sensors data shown in the bottom panel of Fig.~\ref{fig:test1}~(b). The first peak around 0.847~s corresponds to the trigger PMT. About 15~ms later, the second peak appears in particular for pressure sensors P1 and P4, deployed near the two PC covers 1 and 6, which means they collapsed almost simultaneously with their inner PMTs. Then after 100~ms, the third peak takes place, corresponding to acrylic cover 4 and recorded by the nearby pressure sensor P2. Pressure sensor P3 was on top of the central PMT and was found dead after the experiment, presumably due to water leakage, thus the measured pressure magnitude is very small while the time structure is consistent with other sensors. Apparently, the thickness of the cover plays an important role against water pressure and the threshold of the thickness should be between 6.5~mm and 8.5~mm. More important, the 6.5~mm thick acrylic cover resisted water pressure for about 100~ms before collapsing. This period depends on the water pressure and the thickness of the cover. If we can enlarge the inlet area on the stainless steel to let the water go inside faster, the risk of protection cover failure could be reduced, while the primary shockwave due to water flushing in will be increased. Therefore it needs optimization based on the simulation and experiments.

All the underwater experiment results at this stage are summarized in Table.~\ref{tab:test_summary}, including all of the three materials and four production technologies we investigated. They are sorted in thickness on the top of the cover. Because the thickness is not uniform in particular for the covers made of thermoforming and the simulation suggested that the collapse always starts on the top which will be discussed in Sec.~\ref{sec:sim}. The threshold of the implosion of the top cover was thus determined as 7~mm as all 9 samples larger than this thickness succeeded while all but one sample smaller than this thickness failed.

	\begin{table}
	\centering
	\begin{tabular}{ccccc}
		Material & Technology & Thickness on top (mm) & Number of samples & Number of failure \\
		\hline
        PMMA & Injection molding & 11.5 & 5 & 0 \\
        PMMA & Bending & 9.9 & 1 & 0 \\
        PMMA & Casting & 8.5 & 1 & 0 \\
        PMMA & Bending & 8.5 & 1 & 0 \\
        PMMA & Bending & 7.5 & 1 & 0 \\
        PMMA & Bending & 6.8 & 2 & 1 \\
        PMMA & Bending & 6.7 & 1 & 1 \\
        PMMA & Casing & 6.5 & 1 & 1 \\
        PC & Blow molding & 3.0 & 3 & 3 \\
        PETG & Blow molding & 5.0 & 1 & 1 \\
        PETG & Blow molding & 3.0 & 1 & 1 \\
        PETG & Blow molding & 1.5 & 1 & 1 \\
        \hline
	\end{tabular}
	\caption{Summary of underwater PMT protection experiments }
	\label{tab:test_summary}
	\end{table}

\section{Final design and validation}
\label{sec:validation}

\subsection{Simulation study}
\label{sec:sim}

For the thin spherical structure such as the protection cover, buckling is the most common failure mode in case of external pressure. The classical critical buckling pressure $p_b$ of a spherical shell can be expressed as~\cite{buckling}

\begin{equation}\label{eq:bkl}
p_b = \frac{2E}{\sqrt{3(1-\nu^2)}}\frac{t^2}{r^2},
\end{equation}
where E is the tensile modulus, $\nu$ is Poisson's ratio, $t$ and $r$ are the thickness and middle radius of the shell, respectively. The buckling performance for a hollow structure with other shapes is not as good as a sphere and thus needs sophisticated simulation study. In particular, since $r$ is limited by the radius of PMT, with a given material, $t$ is the most sensitive parameter for the buckling pressure and thus needs to be optimized.

\begin{figure}[ht]
\begin{subfigure}{.33\textwidth}
  \centering
  \includegraphics[width=0.9\textwidth]{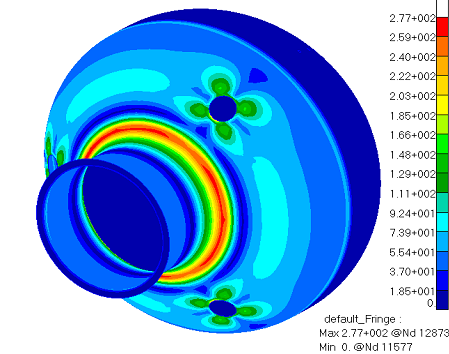}
  \caption{}
  \end{subfigure}
\begin{subfigure}{.33\textwidth}
  \centering
  \includegraphics[width=0.8\textwidth]{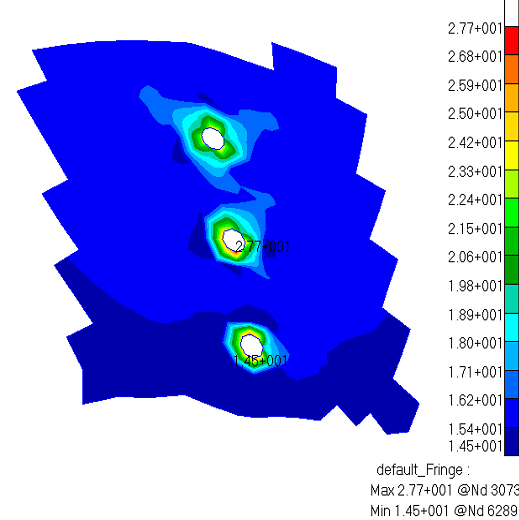}
  \caption{}
  \end{subfigure}
  \begin{subfigure}{.33\textwidth}
  \centering
  \includegraphics[width=0.8\textwidth]{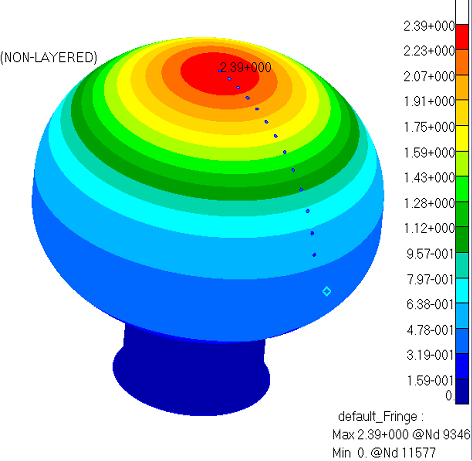}
  \caption{}
  \end{subfigure}
\caption{Finite Element Analysis for a 12~mm thick top cover and 3~mm thick bottom cover with a 0.5~MPa uniform pedestal load. (a) Colors represent tensile stress in the unit of MPa. (b) Tensile stress around the holes on the top cover. (c) Colors represent displacement in the unit of mm.}
\label{fig:sim}
\end{figure}

Finite Element Analysis (FEA) was done taking into account the geometry for both top and bottom covers as two semi-ellipsoids. The inner diameter at the equator was fixed at 512~mm, leaving 2~mm clearance to the PMT. The inner height of the top cover was 186~mm as a baseline, and 14 holes with a diameter between 5~mm and 10~mm were implemented in the simulation. The connection between the top and bottom covers was set as tight. Two independent simulation analyses using different software were done and compared.
Deformation and tensile stress of the protection cover were simulated in the condition of a 0.5~MPa uniform pedestal load, with an example shown in Fig.~\ref{fig:sim}, where the thickness was set as 12~mm for the top cover and 3~mm for the bottom cover.

\begin{figure}[ht]
\begin{subfigure}{.33\textwidth}
  \centering
  \includegraphics[width=1\textwidth]{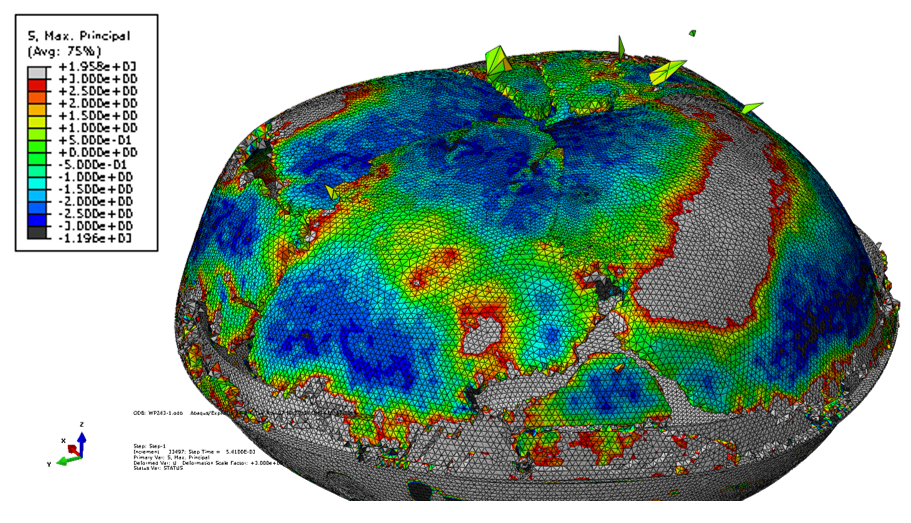}
  \caption{}
  \end{subfigure}
\begin{subfigure}{.33\textwidth}
  \centering
  \includegraphics[width=0.9\textwidth]{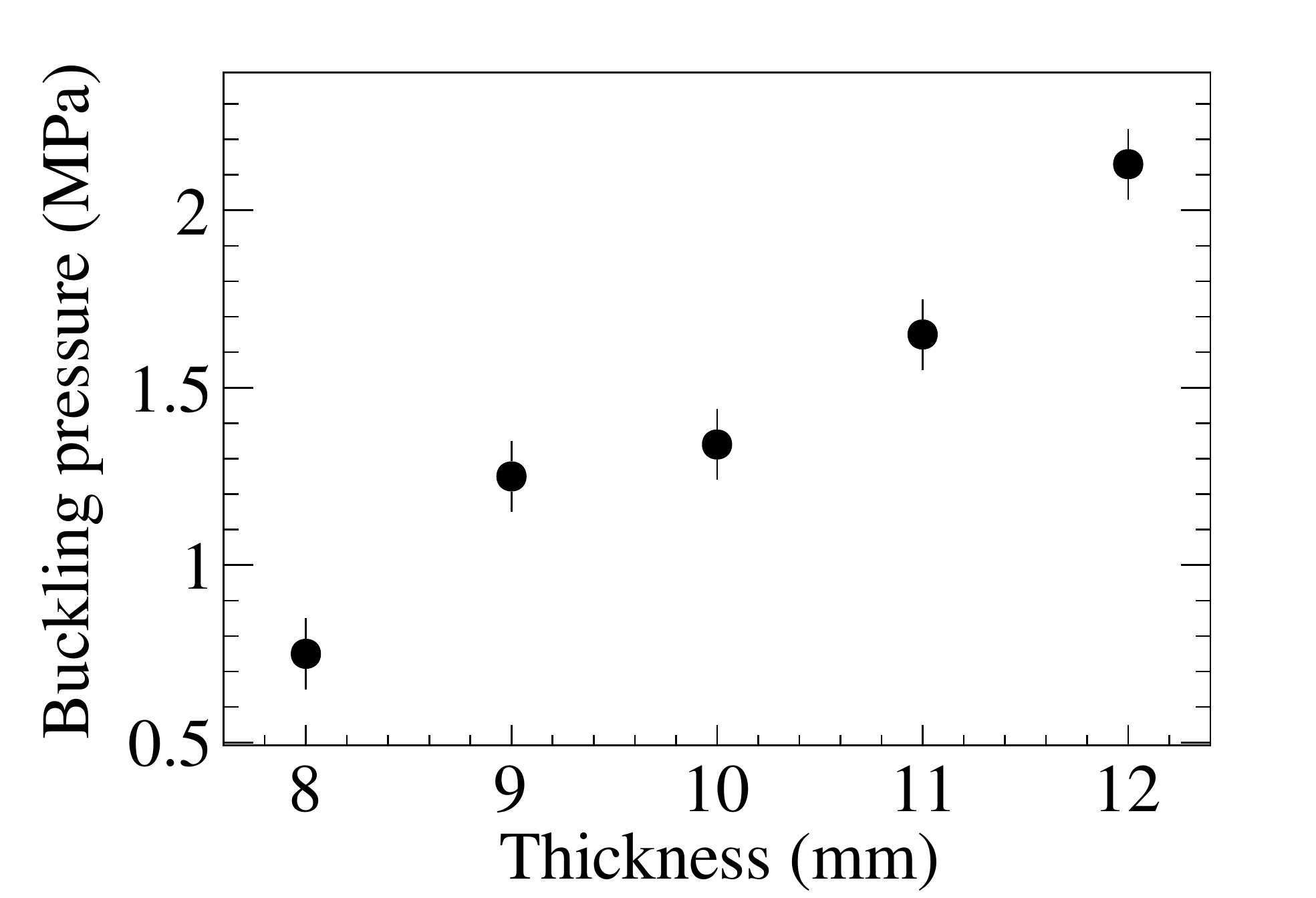}
  \caption{}
  \end{subfigure}
  \begin{subfigure}{.33\textwidth}
  \centering
  \includegraphics[width=0.9\textwidth]{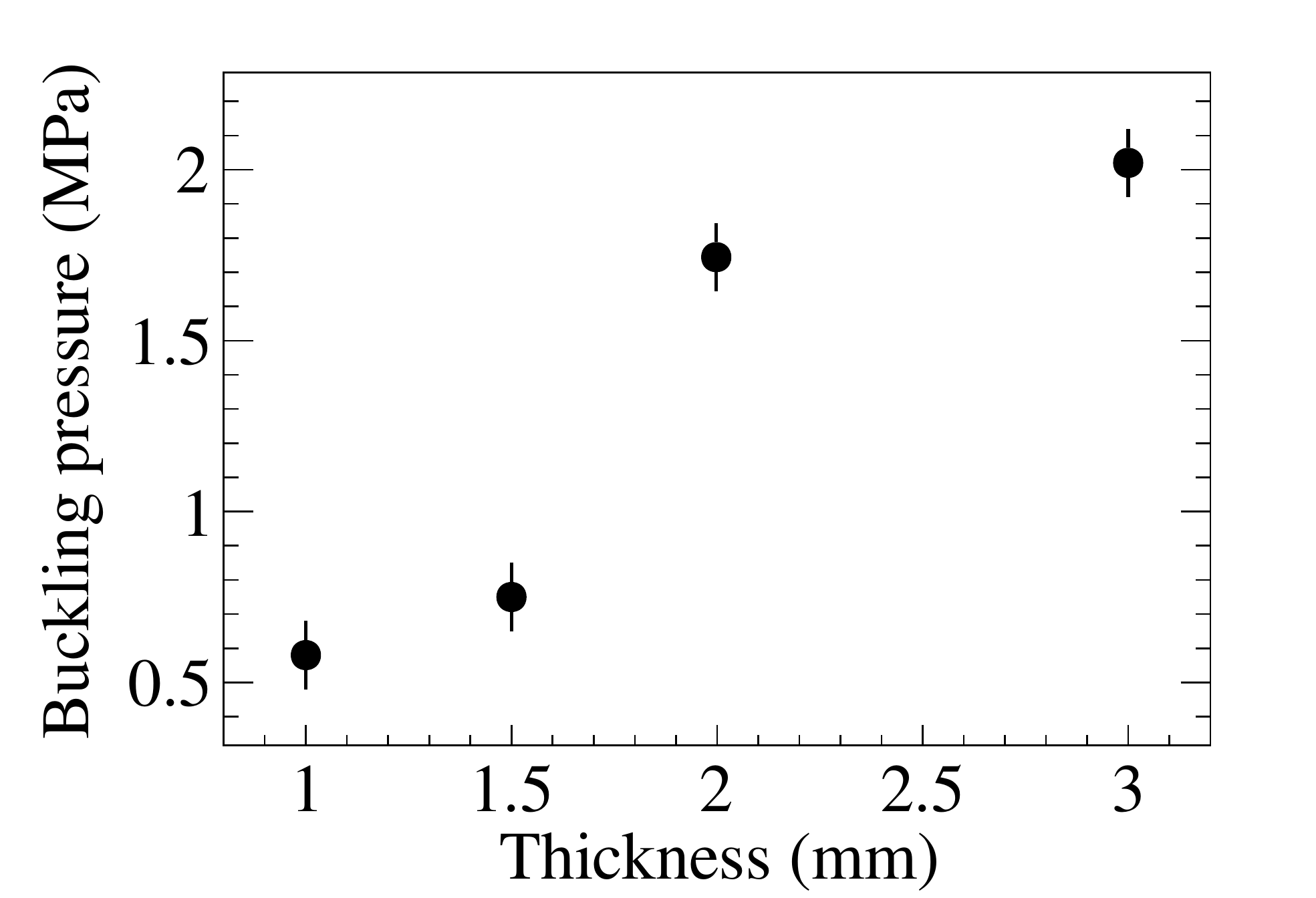}
  \caption{}
  \end{subfigure}
\caption{Finite Element Analysis for buckling failure of protection covers. (a) An example of buckling failure for an 11~mm thick acrylic cover under 2.4~MPa pedestal pressure. Colors represent stress in the protection cover. The deformation was enlarged by a factor of three for better visualization. (b) Critical buckling pressure as a function of acrylic cover thickness. (c) Critical buckling pressure as a function of stainless steel cover thickness. The last data points at 3~mm thickness corresponds to buckling failure of the top cover.}
\label{fig:blsim}
\end{figure}

A buckling failure happens when the deformation is too large that the structure suddenly collapses. A typical buckling failure for an acrylic cover is shown in Fig.~\ref{fig:blsim}~(a), where the buckling starts on the top with many cracks generated. The critical buckling pressure as a function of the acrylic cover thickness is shown in Fig.~\ref{fig:blsim}~(b). The error bars were set as 0.1~MPa, corresponding to the scanning step of the load and the difference between the two analyses. Extrapolating the simulated results to 0.5~MPa buckling pressure with a proportion to the square of the thickness, as suggested in Eq.~\ref{eq:bkl}, gives 6.0~mm, which looks very close to the buckling failure threshold obtained in Table.~\ref{tab:test_summary}.

Since the buckling always starts on the top where there is almost no limitation on the thickness because of detector installation, a non-uniform protection cover was designed with a gradually reduced thickness from top to bottom. Such a design can be easily realized by injection molding. The FEA showed that when the thickness was set as 11~mm on the top and 9~mm at the equator, the buckling pressure was 1.7~MPa, which is three times larger than the largest water pressure in JUNO and the same as a uniform acrylic cover with a thickness of 11~mm. This reduction of 2~mm at the equator gives 1\% increase in the optical coverage. The minimum clearance between two acrylic covers will be 3~mm.

A higher top cover with an inner height 200~mm was also simulated based on the 9-11~mm non-uniform thick design. The buckling pressure was found to be further improved to 2.0~MPa, because it is closer to a sphere. Considering the difficulty of manufacture, larger heights were not considered.

The tensile stress was also studied and found to be much lower than the fracture threshold of acrylic. However, the connection between the top and bottom covers was simplified in the FEA. The stress situation is complicated in the case of a PMT implosion and is hard to be simulated. The study and optimization of the connection will rely on the experiment and will be discussed in Sec.~\ref{sec:connection}.

The critical buckling pressure as a function of stainless steel cover thickness is shown in Fig.~\ref{fig:blsim}~(c), where the top cover was set as 9-11~mm non-uniform thick and 200~mm high. 2~mm thickness was found to be good enough as it resists 1.8~MPa water pressure. Increasing the thickness to 3~mm does not help too much since the top cover starts buckling at 2.0~MPa. At 0.5~MPa pressure, the water flow rate through inlets is about 20~m/s according to Bernoulli’s Principle Formula. Taking into account the resistance of the collapsing PMT glass, if there are four holes 100~mm in diameter on the bottom cover, it will take 10-20~ms for water to fill the PMT volume in case of an implosion. There is negligible impact on the buckling failure with these holes according to the FEA. A cone-shaped bottom cover was also simulated by an independent
analysis and the critical buckling pressure was found to be 40\% lower compared with the semi-ellipsoid one, therefore it was not used for further simulation or prototyping.

\subsection{New prototyping for acrylic covers}

A new injection mold was made according to the final 9-11~mm nonuniform design, with an inner height of 200~mm. Tens of prototypes were produced and found to be solid and smooth without any bubbles inside. However, the steel of the mold was selected to be soft to reduce the cost, and thus the surface can not be polished as good as SPI-A3~\cite{SPI} or better. As a result, the acrylic cover samples look a little fuzzy but it does not have any impact on the mechanical test. The high-transparent acrylic cover has been made with a new mold for mass production, which will not be discussed in this paper.

\subsection{Design of connection}
\label{sec:connection}

\begin{figure}[!hbt]
  \centering
  \includegraphics[width=0.6\textwidth]{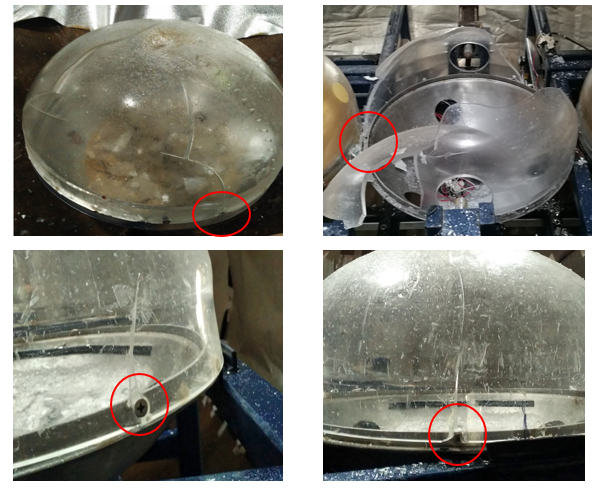}
  \caption{Cracks generated on the acrylic cover in the underwater experiments, all starting from the connection points between top and bottom covers.}
  \label{fig:crack}
\end{figure}

In the past underwater experiments, all of the acrylic covers were found to have some cracks around the connection points, as shown in Fig.~\ref{fig:crack}. There were two processes. When the PMT was imploded, the 0.5~MPa water pressure brought 110~kN force on the acrylic cover, which was supported by six screws with a diameter of several millimeters if the connection was too tight. The contact area between these screws and the acrylic was too small to avoid damage to the acrylic, and there were always cracks starting from the top of the connection holes. After that, water went inside through the big holes on the stainless steel cover, rolled, and pushed the acrylic cover outwards. The force turned into the bottom of the connection holes and sometimes made a notch.

The connection holes on the acrylic cover were optimized as 10~mm in diameter. Six hooks with 6~mm diameter going through the holes pull the acrylic cover down to the flange of the stainless steel cover, one of them shown in the top panel of Fig.~\ref{fig:test22}~(a). With a 1~mm thick gasket surrounding the hole, there is still 2~mm space between the hole and the hook on the top, to avoid direct contact in case of a pedestal water pressure. The distance between the bottom of the hole to the edge of the acrylic cover is 30~mm, three times the diameter of the hole, to reduce the possibility of fracture of the acrylic when the water goes inside.

Another connection structure was designed as a stainless steel ring with an inner diameter a little smaller than the PMT, and fixed with six screws to the stainless steel cover, as shown in Fig.~\ref{fig:test12}~(a). In this case, there are no connection holes on the acrylic cover but this ring will shadow some light thus it was treated as a backup plan. Both of these connections were prototyped and tested in the underwater experiments in Sec.~\ref{sec:exp_val}.

\subsection{Experimental validation}
\label{sec:exp_val}

Experimental validation of the final design was done with a different setup. The central PMT was protected by both the acrylic and the stainless steel cover. One or two PMTs also with protection covers were installed nearby at a minimum clearance of 5~mm, and were used to directly check if there was a chain reaction when the central PMT was imploded. The two implosion plates were moved from the equator to the lower hemisphere and went upwards through the holes on the stainless steel cover to implode the PMT. Another static plate was installed between the acrylic cover and the PMT on the top. The top hole on the acrylic cover was enlarged to 30~mm to let the fixing screw in. In some of the experiments, pressure sensors were also installed to get a quantitative measurement of the shockwave.

\begin{figure}[ht]
\centering
\begin{subfigure}{.42\textwidth}
  \centering
  \includegraphics[width=0.9\textwidth]{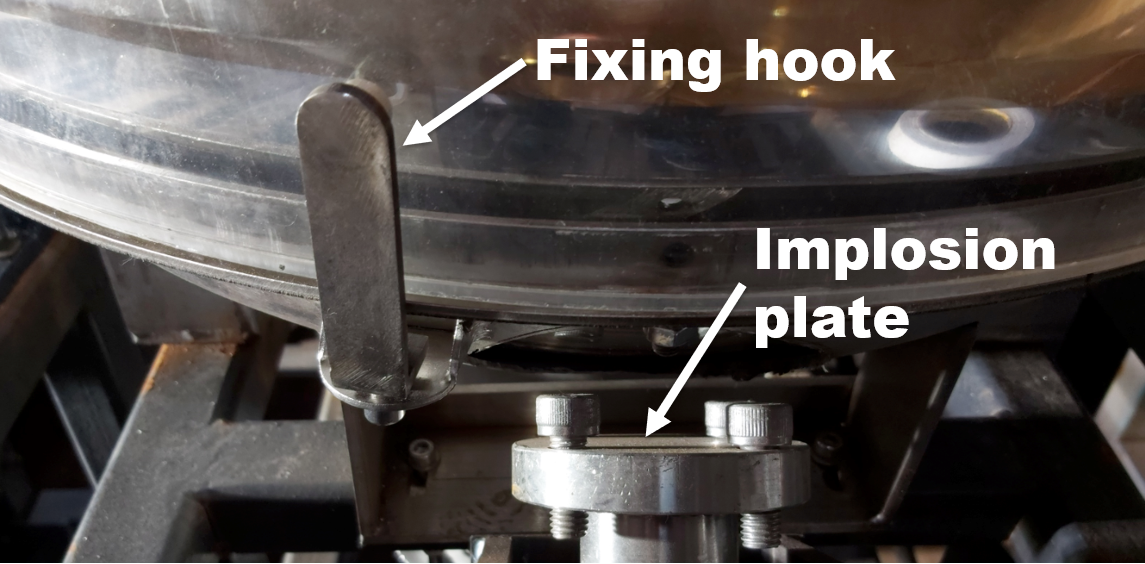}
  \includegraphics[width=0.9\textwidth]{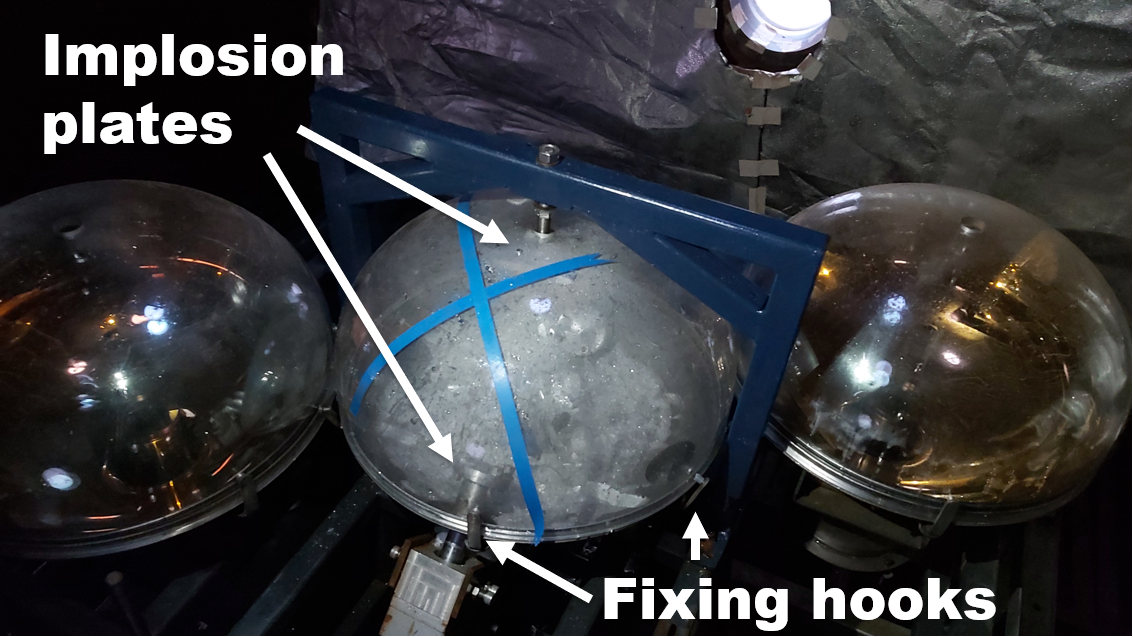}
  \caption{}
\end{subfigure}
\begin{subfigure}{.54\textwidth}
  \centering
  \includegraphics[width=\textwidth]{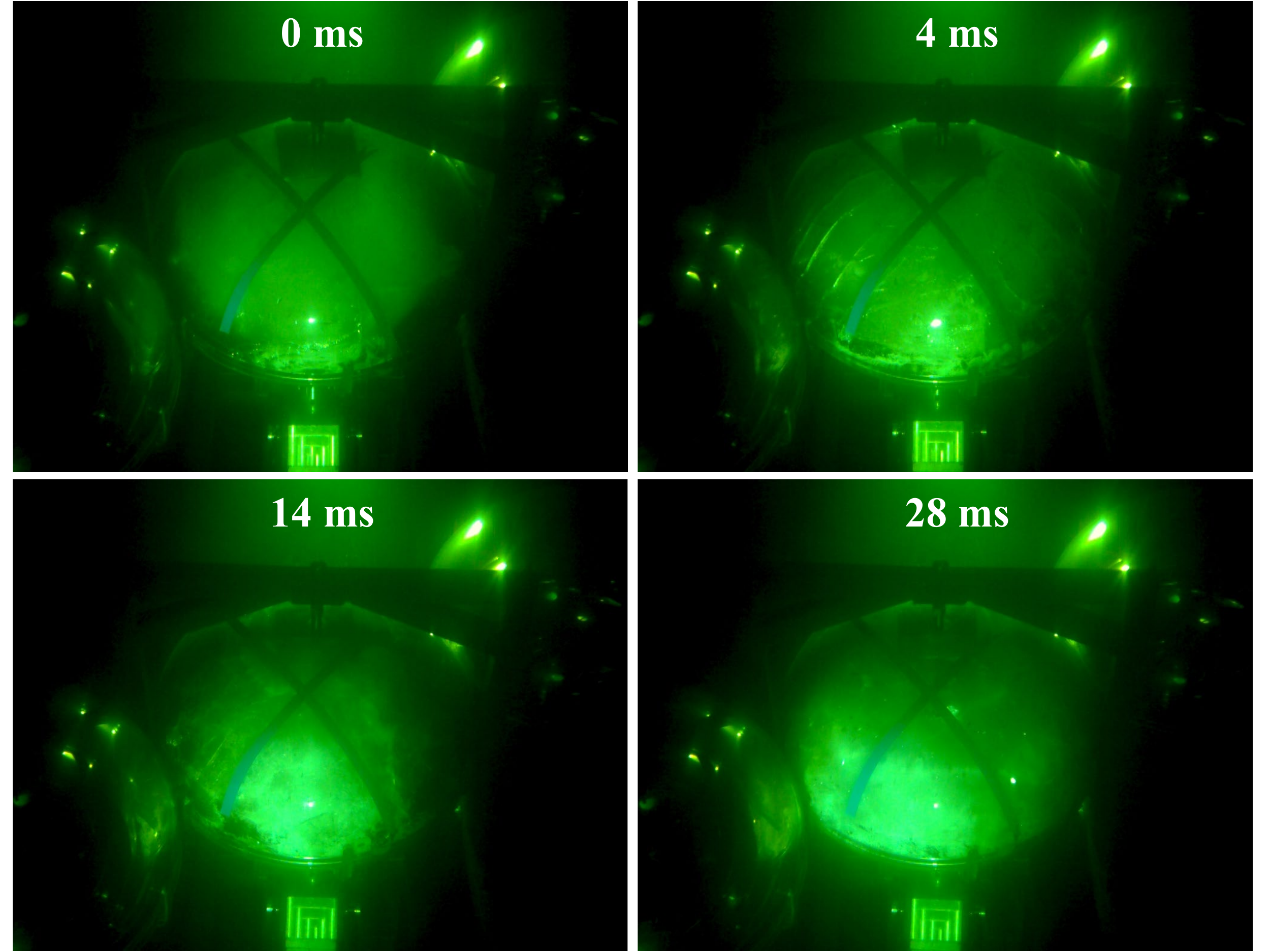}
  \caption{}
\end{subfigure}
\caption{(a) Underwater experiment to validate the final design, with the acrylic cover fixed by six hooks. Two plastic tapes in blue were pasted on the acrylic cover to help the high-speed camera trace possible cracks. The picture in the bottom panel was taken after the implosion of the central PMT. (b) Pictures recorded by the high-speed camera during the implosion of the central PMT.}
\label{fig:test22}
\end{figure}

One of the experiments is shown in Fig.~\ref{fig:test22}~(a). The acrylic cover was made by MGSV and fixed with six hooks. After the implosion, the acrylic cover survived and both neighbor PMTs were safe, proving that there was no chain reaction. Only small cracks were found on the bottom of three of the connection holes, which was completely understood and showed that our optimization of the connection worked well. This experiment was recorded by a high-speed camera at 3,000 frames per second, with four representative frames shown in Fig.~\ref{fig:test22}~(b). At time 0, cracks start on the glass bulb near the implosion plate. 4~ms later, cracks are all over the glass bulb. 14~ms is the middle of PMT collapse. At 28~ms, the volume inside the protection cover is filled with water. The water filling time is larger than the calculation in Sec.~\ref{sec:sim} because two injects were partially blocked by the implosion plates.

\begin{figure}[!hbt]
  \centering
  \begin{subfigure}{.45\textwidth}
  \centering
  \includegraphics[width=0.9\textwidth]{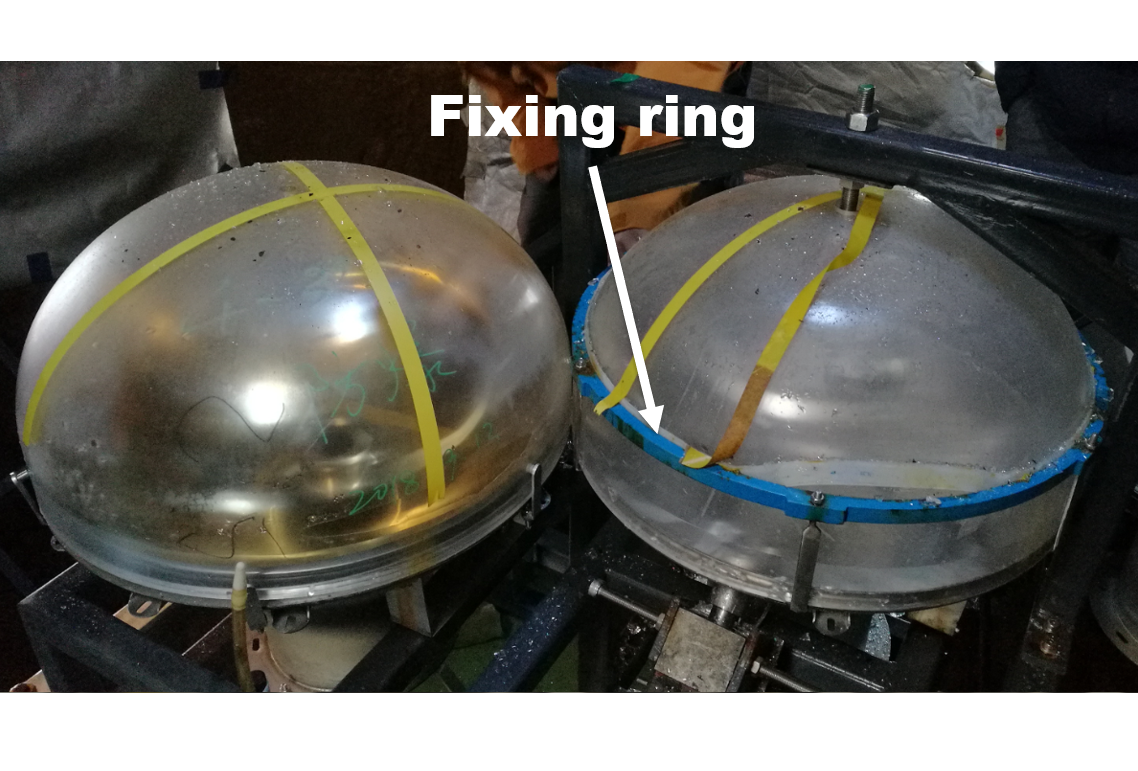}
  \caption{}
\end{subfigure}
\begin{subfigure}{.45\textwidth}
  \centering
  \includegraphics[width=0.9\textwidth]{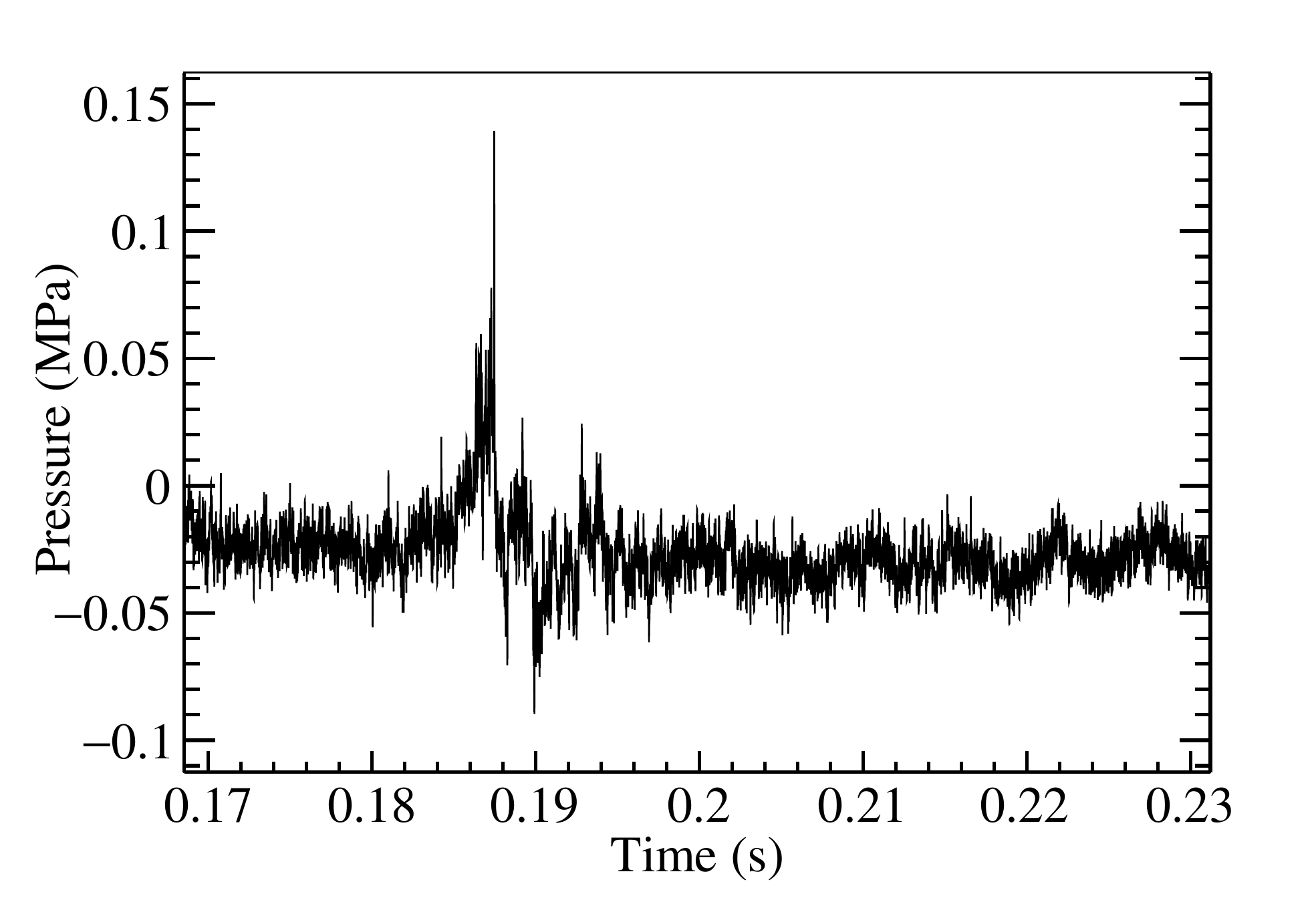}
  \caption{}
\end{subfigure}
  \caption{(a) Underwater experiment to validate the final design, with the acrylic cover of the imploded PMT fixed by the stainless steel ring . (b) Shockwave measured by the pressure sensor. }
  \label{fig:test12}
\end{figure}

In another experiment, IM20 was used to make the acrylic cover and it was fixed with a stainless steel ring as an alternative connection structure, as shown in Fig.~\ref{fig:test12}~(a). It was also successful since the neighbor PMT was undamaged after the implosion of the central PMT. A long going-through crack was found on the acrylic cover below the stainless steel ring, which was also understood, that when the water went inside and pushed the acrylic cover upwards, because of the constraint by the ring, the stress on the acrylic was so large and beyond the fracture strength. A pressure sensor was deployed in this experiment on the top 60~cm far from the center of the imploded PMT. A peak was found with an amplitude of around 0.15~MPa after the baseline subtraction shown in Fig.~\ref{fig:test12}~(b). That was created by the influx of water through the holes on the stainless steel cover. Extrapolating this number to the neighbor PMT glass shell as $L^{-1.2}$ results in 0.37~MPa, a factor of 50 reductions of shockwave compared to the peak pressure in case of no protection.

\section{Summary and perspective}
\label{sec.summary}
A protection system consisting of 20,000 shells with a connection structure was designed for JUNO 20-inch PMTs. Each shell is made of a top acrylic cover and a bottom stainless steel cover, with the shape accommodating both NNVT PMTs and HPK PMTs. High transparency and high mechanical performances are the major requirements for the protection cover. In addition, due to 77.9\% optical coverage in JUNO, almost a factor of two compared to other experiments with a similar scale in the world, dimensions of the protection cover need to be optimized and precisely controlled. PMMA (acrylic), PC, and PET were compared as candidates for the top cover, and acrylic was chosen mainly because of its better transparency. Prototypes of acrylic covers were made by casting, thermoforming, and injection molding. Injection molding was finally selected because of the high production rate and precise control of the dimensions. Thirteen UV-transmission PMMA materials were investigated. IM20 from Evonik and MGSV from Sumitomo were used for prototyping and both of them were validated by underwater experiments. Hygroscopicity was also studied for acrylic and the decrease in mechanical performances due to water absorption was found to be negligible. Prototypes of stainless steel covers were made by rolling and stamping.

More than 10 underwater experiments at 0.5~MPa proceeded which can be divided into two stages. In the first stage, a novel experiment scheme was carried out by imploding multiple protected PMTs using shockwave simultaneously and thus tested a number of protection covers made by different materials or technologies and with different thicknesses at the same time. It largely improved the testing efficiency and reduced the experiment cost. As a result, a thickness of around 7~mm was determined to be the minimum requirement. Simulation studies with FEA were done concerning buckling failure and tensile failure. For the acrylic cover, a nonuniform thickness design was finalized, with 11~mm on the top and 9~mm on the bottom. Two different shapes of the stainless steel cover were compared and a semi-ellipsoid shape was chosen because of better mechanical performance and the thickness was optimized as 2~mm. The critical buckling pressure for the full structure is three times larger than the 50~m water pressure. Four inlet holes with 100~mm diameter were designed on the stainless steel cover to let the water fill in within 10~ms. Connections between the top and bottom covers were also optimized to minimize the risk of acrylic fracture in case of an implosion.

Based on the final design, new prototypes of acrylic covers were produced and validated in the second stage of underwater experiments, where only one protected PMTs was imploded. The neighbor PMTs were safe and the cascade implosion was avoided. Measurement of pressure sensors showed 0.37~MPa peak pressure extrapolated to the neighbor PMTs, which was a factor of 50 reductions of shockwave and far below the PMT implosion threshold. The final design and prototyping of the protection system were proved.

Both the acrylic covers and stainless steel covers are under production. A sampling test shows the transparency of the acrylic cover is larger than 91\% in air, and the manufacturing precision is better than 1~mm. Radioactivity of a few samples was measured and the background contribution from acrylic covers and stainless steel covers is O(10$^{-4}$) and O($10^{-2}$) respectively compared to the PMT itself. We plan to publish another paper on the engineering design, mass production, quality control, and performances once the mass production finishes.

\section{Acknowledgments}
This work was supported by the Strategic Priority Research Program of the Chinese Academy of Sciences, Grant No. XDA10011100. We thank North Night Vision Technology Co., Ltd for providing all PMTs in the underwater experiments. We thank Naval Research Institute, Naval University of Engineering, and Zhejiang University for the simulation studies. Naval Research Institute also provided the high-pressure water tank, high-speed camera, and pressure sensors.

\bibliographystyle{h-physrev5}
\bibliography{references}

\end{document}